\documentclass{elsarticle}

\usepackage{hyperref}
\usepackage{bm}
\usepackage{float}
\usepackage{amsmath}
\usepackage{amsthm}
\usepackage{amsfonts}
\usepackage{graphicx}
\usepackage{comment}

\makeatletter
\def\ps@pprintTitle{%
  \let\@oddhead\@empty
  \let\@evenhead\@empty
  \let\@oddfoot\@empty
  \let\@evenfoot\@oddfoot
}
\makeatother

\usepackage[usenames,dvipsnames]{xcolor}

\newtheorem{remark}{Remark}[section]

\usepackage{color}
\newcommand{\Giacomo}[1]{\textcolor{black}{{#1}}}

\usepackage{setspace}
\begin{document}

\begin{frontmatter}

\title{\Giacomo{An unstructured CD-grid variational formulation for sea ice dynamics}}

\author{Giacomo Capodaglio \fnref{lanlFootnote1}}

\author{Mark R. Petersen \fnref{lanlFootnote2}}

\author{Adrian K. Turner \fnref{lanlFootnote3}}

\author{Andrew F. Roberts \fnref{lanlFootnote3}}

\fntext[lanlFootnote1]{Computational Physics and Methods Group and Center For Nonlinear Studies,
Los Alamos National Laboratory,
e-mail: gcapodaglio@lanl.gov}

\fntext[lanlFootnote2]{Computational Physics and Methods Group,
Los Alamos National Laboratory}

\fntext[lanlFootnote3]{Fluid Dynamics and Solid Mechanics Group,
Los Alamos National Laboratory}

\fntext[LA-UR]{Document approved for unlimited release LA-UR-21-32250.}

\begin{abstract} \Giacomo{For the numerical simulation of earth system models, Arakawa grids are largely employed. A quadrilateral mesh is assumed for their classical definition, and different types of grids are identified depending on the location of the discretized quantities. The B-grid has both velocity components at the center of a cell, the C-grid places the velocity components on the edges in a staggered fashion, and the D-grid is a ninety-degree rotation of a C-grid.}  Historically, B-grid formulations of sea ice dynamics have been dominant because they have matched the grid type used by ocean models. The reason for the grid match is simple – it facilitates penetration of the curl of ice-ocean stress into the deep ocean with minimal numerical diffusivity because sea ice and ocean velocity are co-located.  In recent years, as ocean models have increasingly progressed to C-grids, sea ice models have followed suit on quadrilateral meshes, but few if any implementations of unstructured C-grid sea ice models have been developed. \Giacomo{In this work, we present an unstructured CD-grid type formulation of the elastic-viscous-plastic rheology, where the velocity unknowns are located at the edges  rather than at the vertices, as in the B-grid. Note that the notion of a CD-grid has been recently introduced and assumes that the velocity components are co-located at the edges.}
The mesh cells in our analysis have $n$ sides, with $n$ greater than or equal to four. 
Numerical results are also included to investigate the features of the proposed method.
Our framework of choice is the Model for Prediction Across Scales (MPAS) within E3SM, the climate model of the U.S. Department of Energy, although our approach is general and could be applied to other models as well. While MPAS-Seaice is currently defined on a B-grid, MPAS-Ocean runs on a C-grid, hence interpolation operators are heavily used when coupled simulations are performed. \Giacomo{The discretization introduced here aims at transitioning the dynamics of MPAS-Seaice to a CD-grid mesh, in order to ultimately facilitate improved coupling with MPAS-Ocean and reduce numerical errors associated with this communication.}
\end{abstract}
\end{frontmatter}

\section{Introduction}
Sea ice, saline ice buoyed to the surface of the ocean, plays an important role in the equilibrium of global climate. For instance, its production stimulates ocean overturning, creates a platform for snow cover that in turn greatly increases the planetary albedo of Earth, and forms a marine thermal blanket against the frigid winter polar atmosphere \cite{curry1995sea,kwok2011thinning}.
The ability to model sea ice and predict its state is therefore an important task for climate modelers, and computer simulation is an invaluable tool for this purpose. Many numerical models  of sea ice have been developed since the 1960s, with CICE \cite{hunke2010cice} being perhaps the most widely used in this century owing to its ability to readily exploit parallel computing architectures. CICE is built on a quadrilateral structured mesh and a variational approach is used for the discretization of the divergence of internal ice stress. Within the Model for Prediction Across Scales (MPAS) framework \cite{ringler2010unified,ringler2013multi}, such an approach has been generalized to unstructured grids in the MPAS-Seaice model \cite{turner2021mpas}, using meshes obtained from Voronoi tessellations \cite{ju2011voronoi}. \Giacomo{The Voronoi tessellation is usually called the primal mesh, to which is associated a Delaunay triangulation, referred to as the dual mesh.
Both MPAS-Seaice and CICE are built on an Arakawa B-grid \cite{arakawa1977computational}, where both velocity components are discretized at the center of a cell, whereas the scalar quantities are located on the vertices.} Many existing sea ice models use this kind of staggered grid. \Giacomo{Note that in the case of MPAS-Seaice, it is the dual mesh that is on a B-grid, hence the velocity components are discretized on the vertices of cells of the primal mesh.}
Focusing only on the velocity components, other models that discretize them at the vertices are for instance FESIM \cite{danilov2015finite}, the sea ice component of  FESOM \cite{timmermann2009ocean, wang2014finite}, and LIM \cite{timmermann2005representation}, \Giacomo{although a C-grid placement, where velocity components are discretized at the edge locations in a staggered fashion}, is also available in the latter \cite{bouillon2009elastic}. On the other hand, the ICON-O model features a triangular mesh with a C-grid type staggering and a finite element discretization \cite{korn2017formulation}, where the normal velocity component is defined at the edges of the computational cells.
Interest in C-grid type of methods has grown, in part thanks to a shift in ocean model discretizations from a B-grid to a C-grid. 
\Giacomo{It should be mentioned that for the correct description of the internal sea ice stress, both components of the velocity are needed, and it has been recently observed that on a unstructured triangular grid, it is not sufficient to only consider the normal component of the velocity vector at the edges
\cite{mehlmann2021sea,  danilov2021discretizing} Therefore, the notion of a so called CD-grid has emerged in the aforementioned works, where the components of the velocity vector are co-located at the edges. For a quadrilateral mesh, a CD-grid has twice as many degrees of freedom compared to a B-grid and a C-grid \cite{mehlmann2021simulating}. Hence, for fixed resolution, it is reasonable to expect that the CD-grid will produce lower errors than the B-grid, but it will likely be more computationally expensive.
In \cite{mehlmann2021sea}, a nonconforming Crouzeix-Raviart finite element formulation using a CD-grid was introduced on a triangular mesh. Here, we aim at presenting the mathematical formulation of a variational CD-grid type of approach for unstructured grids with polygonal cells having $n$ sides, with $n \geq 4$. Our focus is on sea ice dynamics, hence we do not discuss the placement of  scalar quantities, and assume that both components of the velocity are discretized at the same edge locations of the mesh. The differences between our approach and that in \cite{mehlmann2021sea} will be discussed in more detail in the rest of the paper.}
 The analysis proposed here is based on the variational strategy for the elastic-viscous-plastic (EVP) rheology  \cite{hunke1997elastic}, laid out in \cite{hunke2002elastic} for a B-grid, and extended to unstructured polygonal meshes for the same type of grid in \cite{turner2021mpas}. Although our method has originated with the MPAS framework in mind, it is general enough to be applied for instance to structured quadrilateral meshes as well.

The paper is organized as follows: in Section \ref{sec:formulation} we lay out the mathematical formulation, highlighting its applicability to a general class of polygonal meshes.
Next, in Section \ref{sec:numRes}, we present a series of test cases in planar and spherical domains to investigate the accuracy and convergence of the proposed method, comparing it with the B-grid formulation currently available in MPAS-Seaice.  Finally, we summarize our findings and discuss future work in Section \ref{sec:concl}.

\section{\Giacomo{Variational formulation on a CD-grid}}\label{sec:formulation}

\Giacomo{In this section, we describe the variational approach applied to an  unstructured polygonal CD-grid.}

\subsection{Preliminaries}
\begin{figure}[!t]
   \centering
   \includegraphics[scale=0.6]{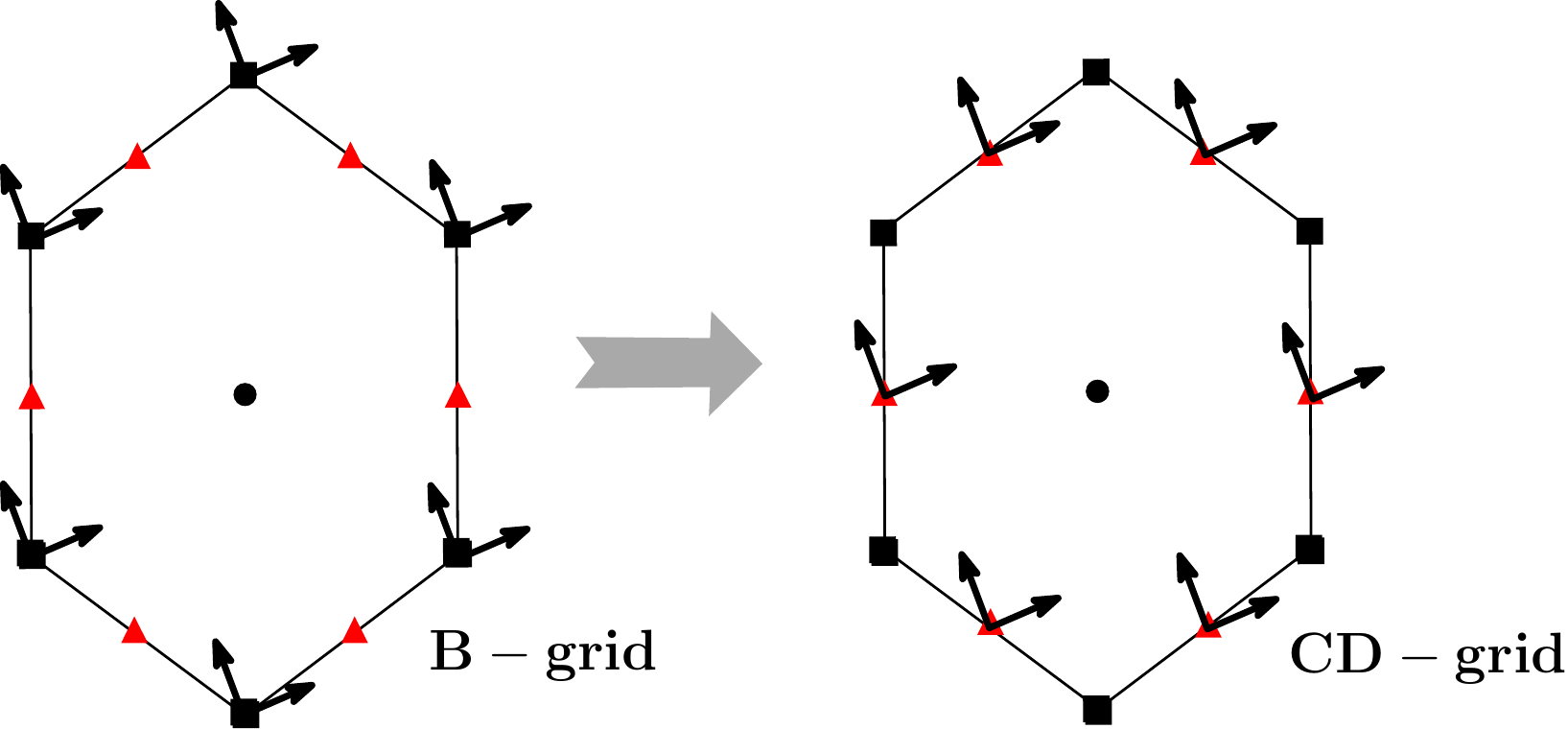}
\caption{\Giacomo{Locations of the reference frames for the B-grid (left), and the CD-grid considered in this work (right). The different symbols identify different discretization points: black squares represent vertex discretization points, red triangles edges, and black dots centers. The meaning of the symbols is the same in Figures \ref{fig:2_bis}, \ref{fig:2}, and \ref{fig:comparison}.}}
   \label{fig:1bis}
\end{figure}

\Giacomo{For the following analysis, we will be considering a spherical domain, hence the coordinate system will be placed on a surface. This means that given a point on the sphere, there is a reference frame lying on the tangent plane to this point, and the problem is two-dimensional on the spherical surface}. 
Let us consider the reduced sea ice momentum equation of Hunke and Dukowikz \cite{hunke1997elastic}

\begin{equation}\label{eq:momentum}
    m \dfrac{\partial \bm{u}}{\partial t} = \nabla \cdot \bm{\sigma} + \bm{\tau}_a + \bm{\tau}_w - \bm k \times m \, f \,\bm{u} - m \, g\, \nabla H_0.
\end{equation}
The left hand side represents the inertial term, with $m$ being the mass of snow and ice per unit area and $\bm{u}$ the sea-ice velocity.
On the right hand side, the first term is the divergence of the ice internal stress $\bm{\sigma}$, $\bm{\tau}_a$ and $\bm{\tau}_w$ are the horizontal stresses due to atmospheric winds and ocean currents respectively, the next term is the Coriolis force and the last takes into account the force coming from the ocean surface tilt. The unit vector $\bm{k}$ is normal to the Earth surface, $f$ is the Coriolis parameter, $g$ is the gravitational acceleration, and $H_0$ is the ocean surface height. \Giacomo{We consider the EVP rheology \cite{hunke1997elastic} that relates the internal stress components $\sigma_{ij}$ with the strain rate tensor components $\dot{\epsilon}_{ij}$ as follows
\begin{equation}
    \dfrac{1}{E}\dfrac{\partial \sigma_{ij}}{\partial t} + \dfrac{1}{2 \eta }\sigma_{ij} + \dfrac{\eta -\zeta}{4 \eta \zeta}\sigma_{kk}\delta_{ij} + \dfrac{P}{4 \zeta}\delta_{ij}=\dot{\epsilon}_{ij},
\end{equation}
where $E$ is Young's modulus, $\eta$ is the shear viscosity, $\zeta$ is the bulk viscosity, $P$ is the pressure and $\delta_{ij}=1$ if $i=j$ and zero otherwise. 
The aim of this section is to discuss how to compute the divergence of the internal stress $\bm{\sigma}$, assuming that both components of the velocity vector are discretized at the edges of the mesh cells, i.e. on a CD-grid, rather than at the vertices, i.e. on a B-grid.}
Our framework of choice is MPAS, of which the
ocean and sea ice components are part of the Energy Exascale Earth System Model (E3SM), developed by the U.S.
Department of Energy, which runs full climate simulations on variable-resolution meshes \cite{golaz2019doe, petersen2019evaluation, caldwell2019doe}.
The present analysis aims at facilitating the coupling of MPAS-Seaice, which currently runs on a B-grid, with MPAS-Ocean, which is on a C-grid instead.
Having the velocities co-located would reduce numerical diffusivity during communication between the two models.
The MPAS codes run on unstructured polygonal meshes obtained from a Voronoi tessellation \cite{ringler2008multiresolution}, normally referred to as the primal mesh, to which is associated a Delaunay triangulation, the dual mesh.\Giacomo{ Cells of the dual are obtained by joining cell centers of the primal, as shown by the dashed triangles in Figure \ref{fig:comparison}.}
For a recent paper on MPAS-type meshes see \cite{hoch2020mpas}. 
In MPAS, the discretization points on the edges are located at the intersection  between line segments joining dual cell centers with primal cell centers.
In MPAS-Ocean, an orthogonal reference frame is placed at every edge of the mesh  with the tangential axis oriented as the edge. Moreover, only one component of the \Giacomo{ocean velocity vector is prognostic}, namely the one that is normal to the edge according to this reference frame. 
On the other hand, for MPAS-Seaice both components of the \Giacomo{sea ice velocity vector are prognostic}, because they are needed \Giacomo{for the computation of the strain rate components, which in turn are required for the computation of the divergence of the internal stress through the constitutive relation.}
We make the assumption that all the orthogonal reference frames at the edge locations are oriented in the same way according to global eastward and northward directions, as is currently in the B-grid formulation of MPAS-Seaice, see Figure \ref{fig:1bis}. 
We refer to the eastward components of the velocity vector $\bm{u}$ as $u$, and denote the northward component of $\bm{u}$ with $v$, hence $\bm{u}=(u,v)$. 
In a similar fashion, a generic vector field $\bm{f}$ will be expressed as $\bm{f}=(f_u, f_v)$, with $f_{u}$ being the component directed eastward and $f_{v}$ the one directed northward.  
This framework for MPAS-Seaice requires that the fields at the edges coming from MPAS-Ocean have to be rotated first, before they can be used as input for the sea ice model. 

%%%%%%%%%%%%
%%%%%%%%%%%%

\subsection{Variational formulation}
We focus on the discretization of the term $\bm{F} := \nabla \cdot \bm{\sigma}$, because the other terms in Eq. \ref{eq:momentum} can be handled in a fairly straightforward way on a CD-grid. \Giacomo{It has been shown in \cite{hunke2002elastic}} that over the entire domain, 
the total work done by the internal stress is equal to the dissipation of mechanical energy:

\begin{equation}\label{eq:initial_stress}
    \int_{\Omega} (\bm{u} \cdot \bm{F}) dA = -\int_{\Omega} ({\sigma}_{11}{ \dot{\epsilon}}_{11}+2{\sigma}_{12}{ \dot{\epsilon}}_{12} + {\sigma}_{22}{ \dot{\epsilon}}_{22}) dA,
\end{equation}
where $\bm{\sigma}$ is the internal stress tensor and $\bm{\dot \epsilon}$ is the strain rate tensor, \Giacomo{ which is computed as a function of velocity as follows
\begin{equation}\label{eq:strain_rate_initial}
\dot{\epsilon}_{11} = \dfrac{\partial u}{\partial x} -\dfrac{v \tan(\lambda)}{r}, 
\quad \dot{\epsilon}_{12} = \frac{1}{2}\Big(\dfrac{\partial u}{\partial y} + \dfrac{\partial v}{\partial x} + \dfrac{u \tan(\lambda)}{r}\Big),
\quad \dot{\epsilon}_{22} = \dfrac{\partial v}{\partial y}.
\end{equation}
In the above equation, $\lambda$ is the latitude and $r$ the spherical domain's radius. The terms involving the latitude are called metric terms \cite{hunke2002elastic}, and take into account the curvature of the computational grid when a spherical domain is considered. When the domain is planar these terms are neglected since no grid curvature is present.} 
Note that, for simplicity, boundary terms are ignored in Eq. \eqref{eq:initial_stress}.
 \begin{remark}
 To avoid tedious notation, in this analysis we assume we are dealing with \Giacomo{a spherical domain without any continents}, hence no boundaries are present on the computational grid. The results of the analysis do not change in case continents or boundaries are present, and the treatment of coastal boundary conditions follows either a Dirichlet or a Neumann approach. 
 \end{remark}
\Giacomo{Following the variational approach from \cite{turner2021mpas}, we define the function
\begin{equation}
    \Theta(x,y,u,v,\sigma_{11}, \sigma_{12}, \sigma_{22}) := (\bm{u} \cdot \bm{F}) +  ({\sigma}_{11}{ \dot{\epsilon}}_{11}+2{\sigma}_{12}{ \dot{\epsilon}}_{12} + {\sigma}_{22}{ \dot{\epsilon}}_{22}).
\end{equation}
 Then, Eq. \eqref{eq:initial_stress} can be reformulated as
\begin{align}\label{eq:int_functional}
   \mathcal{I}(\bm{u}, \bm{\sigma}) := \int_{\Omega}  \Theta(x,y,u,v,\sigma_{11}, \sigma_{12}, \sigma_{22}) dA = 0.
\end{align}
The Euler-Lagrange equations associated with the functional in Eq. \eqref{eq:int_functional} are given by
\begin{equation}\label{eq:EL_eqs}
\begin{cases}
    \dfrac{\partial \Theta} {\partial u} - \dfrac{\partial } {\partial x} \Big( \dfrac{\partial \Theta} {\partial u_x }\Big) - \dfrac{\partial } {\partial y} \Big( \dfrac{\partial \Theta} {\partial u_y }\Big) = 0,\\  \dfrac{\partial \Theta} {\partial v} - \dfrac{\partial } {\partial x} \Big( \dfrac{\partial \Theta} {\partial v_x }\Big) - \dfrac{\partial } {\partial y} \Big( \dfrac{\partial \Theta} {\partial v_y }\Big) = 0,  \\
        \dfrac{\partial \Theta} {\partial \sigma_{ij}} - \dfrac{\partial } {\partial x} \Big( \dfrac{\partial \Theta} {\partial {\sigma_{ij}}_x }\Big) - \dfrac{\partial } {\partial y} \Big( \dfrac{\partial \Theta} {\partial {\sigma_{ij}}_y }\Big) = 0, \quad (i,j) \in \{(1,1), (1,2), (2,2)\}.
    \end{cases}
\end{equation}
Considering that there is no explicit dependence of $\Theta$ on the partial derivatives of the velocity components, the system of five equations in Eq. \eqref{eq:EL_eqs} reduces to
\begin{equation}\label{eq:EL_eqs_short}
\begin{cases}
    \dfrac{\partial \Theta} {\partial u}   = 0,\\  \dfrac{\partial \Theta} {\partial v}   = 0,  \\
        \dfrac{\partial \Theta} {\partial \sigma_{ij}} - \dfrac{\partial } {\partial x} \Big( \dfrac{\partial \Theta} {\partial {\sigma_{ij}}_x }\Big) - \dfrac{\partial } {\partial y} \Big( \dfrac{\partial \Theta} {\partial {\sigma_{ij}}_y }\Big) = 0, \quad (i,j) \in \{(1,1), (1,2), (2,2)\}.
    \end{cases}
\end{equation}
In the planar case, in can be shown that the three equations above involving the stress components $\sigma_{ij}$ recover the equations in Eq. \eqref{eq:strain_rate_initial}. 
%%%%%%%%%%%%
\begin{figure}[!t]
   \centering
     \includegraphics[scale=0.7]{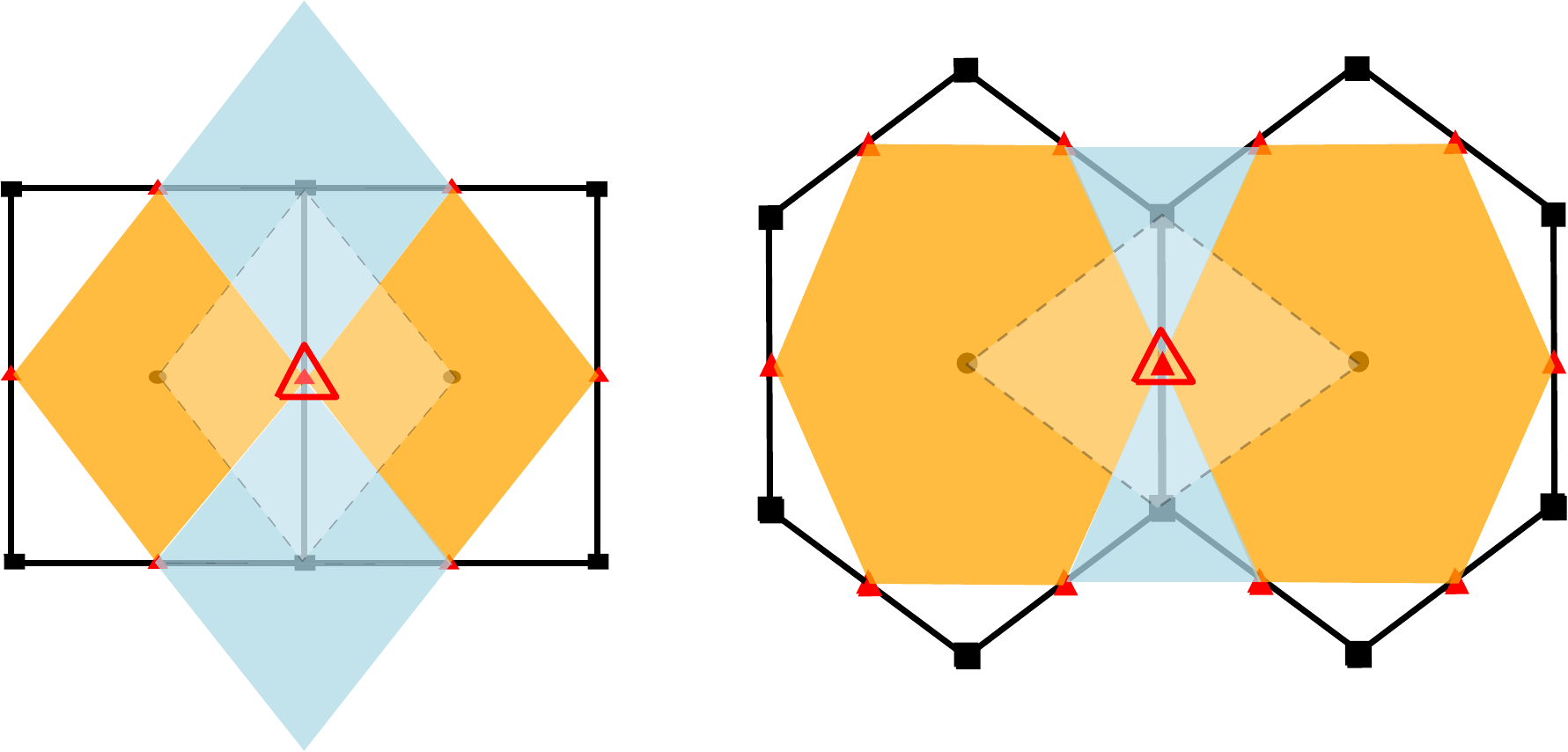}
\caption{\Giacomo{Representation of the control volume $\mathcal{V}_e$ in Eq. \eqref{eq:integrated_eqs}. For any edge $e$ it is the union of the two polygons $\widehat{V}^1_e$ and $\widehat{V}^2_e$ centered around cell centers (orange) and the two triangles $T^1_e$ and $T_e^2$,  centered around vertices (blue-gray). The transparent white region is the diamond-shaped polygon $P_e$ used for the piece-wise constant expansion in Eq. \eqref{eq:const_exp}. Left: quadrilateral mesh. Right: Voronoi tessellation.
}}
   \label{fig:2_bis}
\end{figure}
%%%%%%%%%%%%
Next, for any edge $e$, we define a collection of four polygons $\widehat{V}_e^1, \widehat{V}_e^2, T_e^1, T_e^2$ as follows: for $i\in \{1,2\}$, the polygons $\widehat{V}_e^i$ are obtained by joining the edge points of the primal cell $V^i_e$ that owns $e$ (see the orange shapes in Figure \ref{fig:2_bis}), whereas the polygons $T_e^i$ are obtained by joining with the point on
$e$ the edge points on those edges $e'$ that share a vertex with $e$ and that belong to one of the cells that owns e
(see the blue-gray shapes in Figure \ref{fig:2_bis}). Note that these four polygons only overlap on their boundaries.
\begin{remark}
In case of a quadrilateral mesh as in Figure \ref{fig:2_bis} left, all the $\widehat{V}_e^1, \widehat{V}_e^2, T_e^1, T_e^2$ are quadrilaterals. In case of a Voronoi tessellation as in Figure \ref{fig:2_bis} right, the $\widehat{V}_e^1, \widehat{V}_e^2$ the are polygons with five or more sides, while the $T_e^1, T_e^2$ are triangles.
\end{remark}
%%%%%%%%%%%%
\begin{figure}[!t]
   \centering
   \includegraphics[scale=0.58]{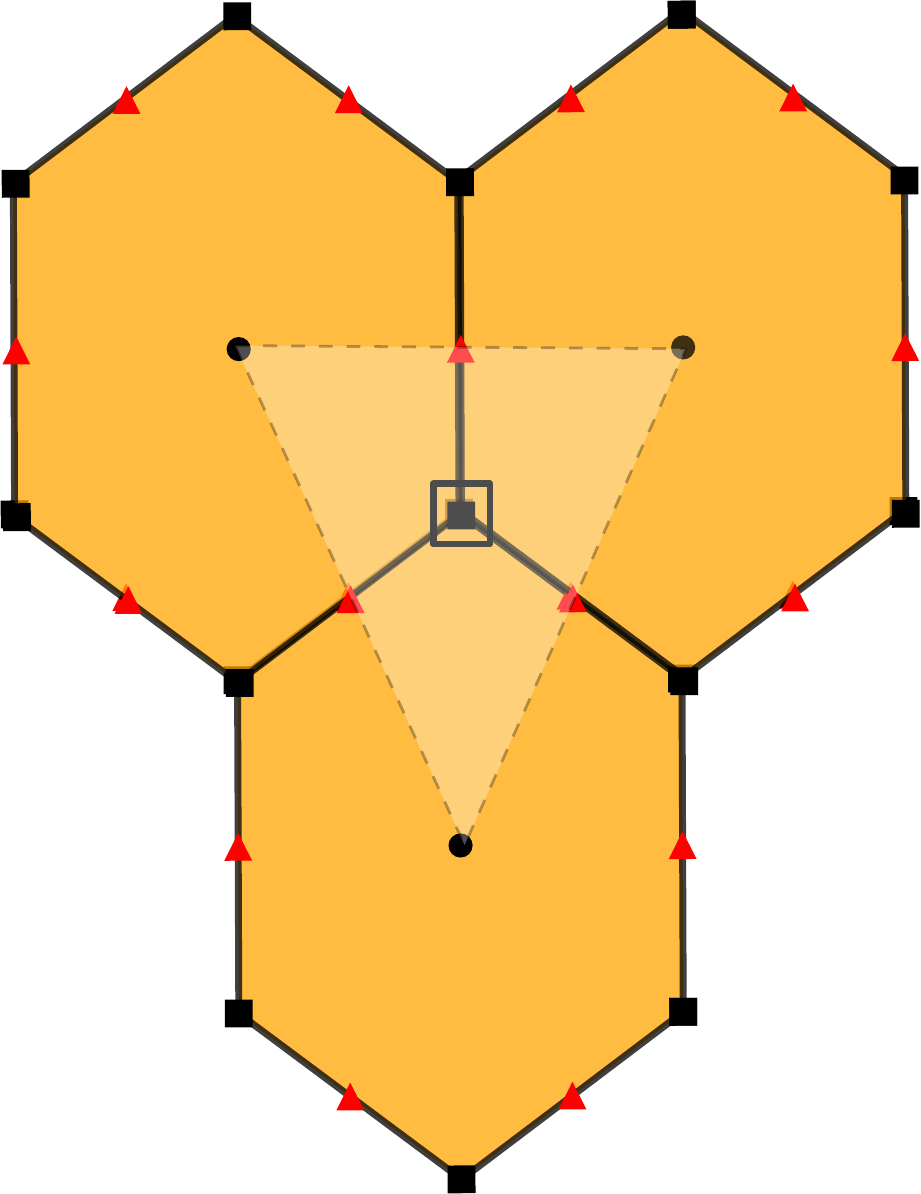}\quad
     \includegraphics[scale=0.58]{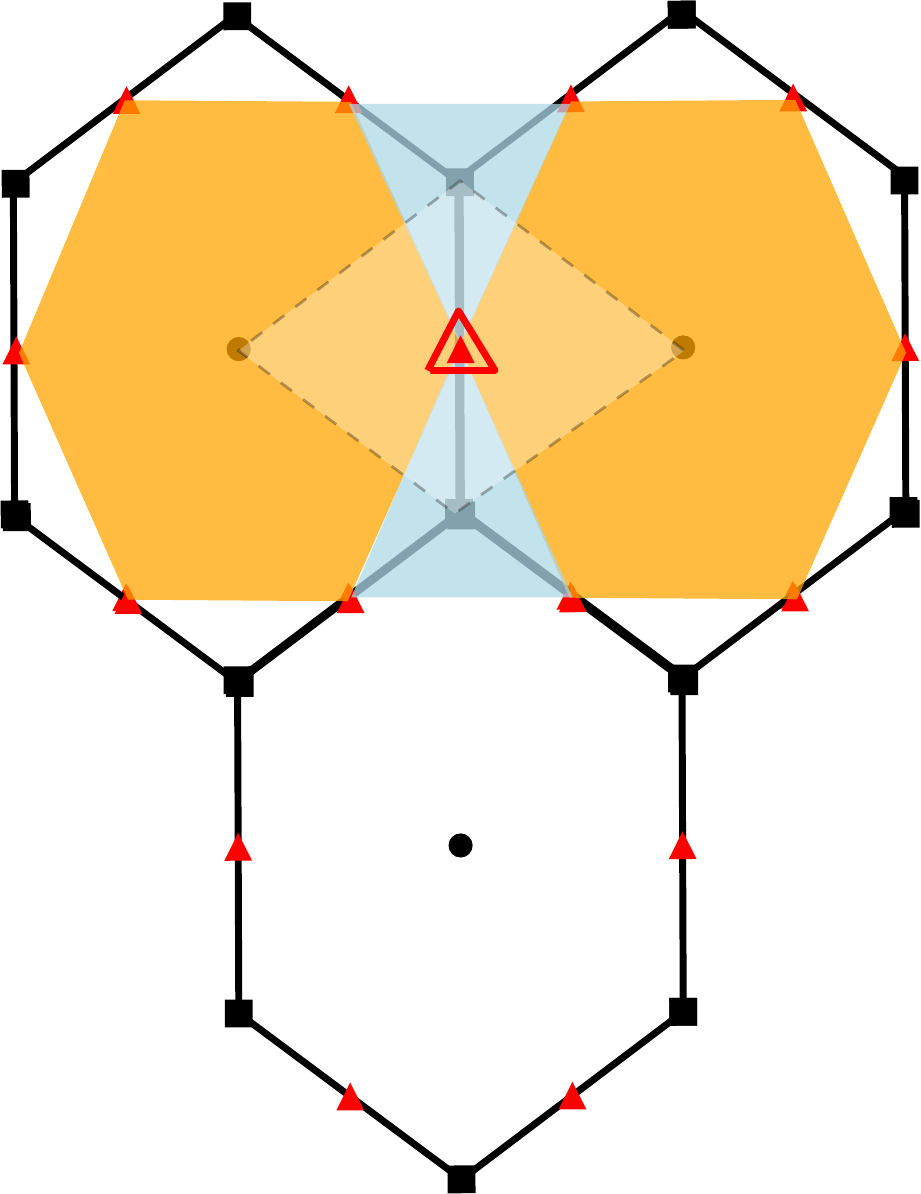}
\caption{
\Giacomo{Comparison of the control volumes used in Eq. \eqref{eq:integrated_eqs}.
Left: B-grid. The velocity at a vertex (black double square) receives contributions from the three cells that own it. Right: CD-grid. The velocity at an edge (red double triangle) receives contributions from the two polygons and two triangles that own it.}}
   \label{fig:2}
\end{figure}
%%%%%%%%%%%%
For any edge $e$, we define the following control volume
\begin{equation}\label{eq:control_vol}
    \mathcal{V}_e = \widehat{V}_e^1 \cup  \widehat{V}_e^2 \cup T_e^1 \cup T_e^2.
\end{equation}
In the existing formulation on MPAS-Seaice that relies on a B-grid, for a given vertex the control volume is given by the union of the primal cells that own that vertex, see Figure \ref{fig:2} (left) for the case of a Voronoi mesh.
 We then integrate the first two equations in Eq.\eqref{eq:EL_eqs_short} over  $\mathcal{V}_e$ 
\begin{equation}
\begin{aligned}\label{eq:integrated_eqs}
         \int_{\mathcal{V}_e} \dfrac{\partial}{\partial u}\Big(\bm{u} \cdot \bm{F}\Big) \,dA = -\int_{\mathcal{V}_e} \dfrac{\partial}{\partial u}\Big({\sigma}_{11}{ \dot{\epsilon}}_{11}+2{\sigma}_{12}{ \dot{\epsilon}}_{12} + {\sigma}_{22}{ \dot{\epsilon}}_{22}\Big) \,dA, \\
             \int_{\mathcal{V}_e} \dfrac{\partial}{\partial v}\Big(\bm{u} \cdot \bm{F}\Big) \, dA = -\int_{\mathcal{V}_e} \dfrac{\partial}{\partial v}\Big({\sigma}_{11}{ \dot{\epsilon}}_{11}+2{\sigma}_{12}{ \dot{\epsilon}}_{12} + {\sigma}_{22}{ \dot{\epsilon}}_{22}\Big) \, dA.
\end{aligned}
\end{equation}
For ease of notation, let
\begin{equation}
\begin{aligned}
    &D_{u,e}^1 := -\int_{\mathcal{V}_e} \dfrac{\partial}{\partial u}\Big({\sigma}_{11}{ \dot{\epsilon}}_{11}\Big)\,dA, \qquad D_{u,e}^2:=   -\int_{\mathcal{V}_e} \dfrac{\partial}{\partial u}\Big(2{\sigma}_{12}{ \dot{\epsilon}}_{12}\Big) \,dA, \qquad  D_{u,e}^3:=  -\int_{\mathcal{V}_e} \dfrac{\partial}{\partial u}\Big({\sigma}_{22}{ \dot{\epsilon}}_{22}\Big)\,dA,  \\
        &D_{v,e}^1 := -\int_{\mathcal{V}_e} \dfrac{\partial}{\partial v}\Big({\sigma}_{11}{ \dot{\epsilon}}_{11}\Big)\,dA, \qquad D_{v,e}^2:=   -\int_{\mathcal{V}_e} \dfrac{\partial}{\partial v}\Big(2{\sigma}_{12}{ \dot{\epsilon}}_{12}\Big) \,dA, \qquad  D_{v,e}^3:=  -\int_{\mathcal{V}_e} \dfrac{\partial}{\partial v}\Big({\sigma}_{22}{ \dot{\epsilon}}_{22}\Big)\,dA, \\
        & \hspace{4.5cm} D_{u,e}:=D_{u,e}^1+D_{u,e}^2+D_{u,e}^3, \qquad D_v:=D_{v,e}^1+D_{v,e}^2+D_{v,e}^3.
    \label{eq:theDs}
    \end{aligned}
\end{equation}
Then Eq. \eqref{eq:integrated_eqs} becomes
\begin{equation}
\begin{aligned}\label{eq:integrated_eqs2}
         \int_{\mathcal{V}_e} \dfrac{\partial}{\partial u}\Big(u\,F_u + v\,F_v\Big) \,dA = D_{u,e}, \qquad 
             \int_{\mathcal{V}_e} \dfrac{\partial}{\partial v}\Big(u\,F_u + v\,F_v\Big) \, dA = D_{v,e}.
\end{aligned}
\end{equation}
\subsubsection{Standard approach} \label{sec:standard}
For the computation of the integrals on the left hand side of Eq. \eqref{eq:integrated_eqs2}, we use a piece-wise constant approximation of the fields $\bm{u}$ and $\bm{F}$ subordinate to a certain cover of the domain $\Omega$.  This approach was used in the original variational formulation \cite{hunke2002elastic} but it has been later improved in \cite{turner2021mpas}, as it will be shown in Section \ref{sec:alternative}.  For the B-grid, the cover is given by the dual triangles associated with the vertices of the mesh, see the transparent triangles in Figure \ref{fig:2} (left). For the CD-grid, this cover uses diamond-shaped polygons $P_e$ obtained by joining the vertices that are
the end points of the edge $e$ with the cell centers of the primal cells $V_e^1$ and $V_e^2$ that own $e$, see the transparent shapes in Figure \ref{fig:2_bis} and Figure \ref{fig:2} (right).
%%%%%%%%%%%%
%\begin{figure}[!t]
%   \centering
%   \includegraphics[scale=0.55]{figures/pic1_quad_hex.pdf}
%\caption{Decomposition of the computational domain, where computational cells $V_c$ are delimited by solid lines. The diamond-shaped polygons $P_e$ as in Eq. \eqref{eq:decomp1}, are delimited by dashed lines and are each uniquely associated with an edge location (red triangle marker). Left: quadrilateral mesh. Right: Voronoi tessellation.}
%   \label{fig:1}
%\end{figure}
%%%%%%%%%%%%
Hence, for the computation of the left hand side integrals in Eq. \eqref{eq:integrated_eqs2}, the velocity and stress divergence fields are expanded as
\begin{align}\label{eq:const_exp}
    u = \sum_{e=1}^{N_e} u_e \,\chi_{P_e}(x,y), \qquad F_u = \sum_{e=1}^{N_e} F_{ue}\, \chi_{P_e}(x,y), \qquad v =   \sum_{e=1}^{N_e} v_e \, \chi_{P_e}(x,y), \qquad F_v =  \sum_{e=1}^{N_e} F_{ve} \,\chi_{P_e}(x,y),
\end{align}
where $N_e$ is the total number of edges of the mesh, $(u_e,v_e) := \bm{u}(x_e,.y_e)$ and $(F_{ue},F_{ve}) := \bm{F}(x_e,.y_e)$ are the values of the fields at the edge locations, and $\chi_{P_e}$ is the characteristic function of the set $P_e$.
Note that Eq. \eqref{eq:const_exp} is only used within integrals and so the values of the fields at the boundary of the $P_e$ sets do not matter. With the expansion in Eq. \eqref{eq:const_exp} and approximating the derivatives $\partial/\partial u$ and $\partial/\partial v$ with $\partial/\partial u_e$ and $\partial/\partial v_e$, the integrals on the left hand side of \eqref{eq:integrated_eqs2} become
\begin{equation}
    \begin{aligned}
            \int_{\mathcal{V}_e} \dfrac{\partial}{\partial u}\Big(u\,F_u + v\,F_v\Big) \,dA = \int_{\mathcal{V}_e} \dfrac{\partial}{\partial u_e}\Big(\sum_{i=1}^{N_e} u_i \,\chi_{P_i} \, \sum_{j=1}^{N_e} F_{uj} \,\chi_{P_j} + \sum_{k=1}^{N_e} v_k \,\chi_{P_k} \, \sum_{l=1}^{N_e} F_{vl} \,\chi_{P_l}\Big) \,dA, \\
            \int_{\mathcal{V}_e} \dfrac{\partial}{\partial v}\Big(u\,F_u + v\,F_v\Big) \,dA =   \int_{\mathcal{V}_e} \dfrac{\partial}{\partial v_e}\Big(\sum_{i=1}^{N_e} u_i \,\chi_{P_i} \, \sum_{j=1}^{N_e} F_{uj} \,\chi_{P_j} + \sum_{k=1}^{N_e} v_k \,\chi_{P_k} \, \sum_{l=1}^{N_e} F_{vl} \,\chi_{P_l}\Big) \,dA.
    \end{aligned}
\end{equation}
Thanks to the Euler-Lagrange framework, the dependence of $\bm{F}$ on $\bm{u}$ can be neglected, since the velocity and the stress are independent parameters of the functional in Eq. \eqref{eq:int_functional}. Hence, the above equations give
\begin{equation}
    \begin{aligned}
            \int_{\mathcal{V}_e} \dfrac{\partial}{\partial u}\Big(u\,F_u + v\,F_v\Big) \,dA =   \int_{\mathcal{V}_e}   \sum_{j=1}^{N_e} F_{uj} \,\chi_{P_j} \chi_{P_e}   \,dA = \int_{P_e} F_{ue} \, dA= F_{ue} A_{P_e} \\
            \int_{\mathcal{V}_e} \dfrac{\partial}{\partial v}\Big(u\,F_u + v\,F_v\Big) \,dA =  \int_{\mathcal{V}_e} \sum_{l=1}^{N_e} F_{vl} \,\chi_{P_l} \chi_{P_e}  \,dA = \int_{P_e} F_{ve} \, dA= F_{ve} A_{P_e}.
    \end{aligned}
\end{equation}
The second to last equality follows because the $P_e$ sets only overlap on their boundaries and $\mathcal{V}_e \cap P_e = P_e$. The quantity $A_{P_e}$ is the area of the diamond-shaped polygon $P_e$.
With these computations, the equations in Eq. \eqref{eq:integrated_eqs2} give
\begin{equation}\label{eq:integrated_eqs3}
    F_{ue}  = \dfrac{D_{u,e}}{A_{P_e}}, \qquad F_{ve}  = \dfrac{D_{v,e}}{A_{P_e}}.
\end{equation}
Let's now continue by making the terms in Eq. \eqref{eq:theDs} explicit. For simplicity, we consider only the $D_{u,e}^i$ (with $i=1,2,3$) since the derivation is similar for the $D_{v,e}^i$:
\begin{equation}\label{eq:D1}
    D_{u,e}^1 = -\int_{\mathcal{V}_e} \dfrac{\partial}{\partial u}\Big( \sigma_{11} \Big[ \dfrac{\partial u} {\partial x} - v \, C_1(r) \tan(\lambda)\Big]\Big) dA,
\end{equation}
\begin{equation}\label{eq:D2}
    D_{u,e}^2 =-\int_{\mathcal{V}_e} \dfrac{\partial}{\partial u}\Big( \sigma_{12} \Big[ \dfrac{\partial u} {\partial y} + \dfrac{\partial v}{\partial x} + u \, C_2(r)\tan(\lambda)\Big]\Big) dA,
\end{equation}
\begin{equation}\label{eq:D3}
    D_{u,e}^3 =-\int_{\mathcal{V}_e} \dfrac{\partial}{\partial u}\Big(\sigma_{22}\Big[ \dfrac{\partial v}{\partial y} + v \, C_3(r) \tan(\lambda)\Big]\Big)dA,
\end{equation}
where $C_i(r)$, $i=1,2,3$ is either identically zero, if the domain is planar, or equal to $b_i/r$ if the domain is spherical, where $b_i$ is a non-negative number that does not depend on $r$. Recall that $r$ is the radius of the spherical domain and $\lambda$ represents the latitude. The terms that include $\tan(\lambda)$ are the metric terms for the divergence of the stress. For any edge $e$, by definition \eqref{eq:control_vol} it holds that 
\begin{equation}
    \int_{\mathcal{V}_e} \, dA = \int_{\widehat{V}_e^1} \,dA + \int_{\widehat{V}_e^2} \, dA + \int_{T_e^1} \,dA +   \int_{T_e^2} \,dA.
\end{equation}
Therefore, for the computation of the integrals on the right hand side of \eqref{eq:integrated_eqs2}, we approximate the value of any of the functions $u$, $v$ and $\sigma_{ij}$ with a basis expansion, which is different depending on whether we are in $\widehat{V}_e^i$ or $T_e^i$.
Namely, for any of the functions $u$, $v$ or $\sigma_{ij},$ we approximate their value at any point $(x,y)$ in $T_e^i$ as the linear combination of basis functions centered at the vertices of $T_e^i$
 \begin{equation}\label{eq:basis1}
     f(x,y) = \sum_{j=1}^{n_t} f_{t_j} \mathcal{L}_{t_j}(x,y),
 \end{equation}
 where $n_t$ is the number of vertices of $T_e^i$, $f$ is any of the functions mentioned above, $f_{t_j}$ is the value of the function at the $j$-th vertex of $T_e^i$ (which is an edge location) and $\mathcal{L}_{t_j}$ is the basis function associated with vertex $t_j$.
For ease of notation, the explicit dependence on $e$ is not included in Eq. \eqref{eq:basis1}. We choose basis functions $\mathcal{L}_{t_j}$ for which
       \begin{equation}\label{properties}
 \begin{aligned}
 \begin{cases}
        \sum_{j=1}^{n_t} \mathcal{L}_{t_j}(x,y)&=1 \, \mbox{  for any } (x,y) \in T_e^i \\
        \mathcal{L}_{t_j}(x_{t_k},y_{t_k}) &= \delta_{jk}
        \end{cases},
 \end{aligned}
 \end{equation}
 where $(x_{t_k},y_{t_k})$ is the location of the $k$-th vertex of $T_e^i$ and $\delta_{jk}$ is Kronecker's delta.
 Because of these features, the function $f$ is approximated with a function similar to a finite element nodal interpolator.
 The same approach is used for any polygon $\widehat{V}_e^i$, with basis functions centered at the vertices of $\widehat{V}_e^i$, which are the edge points of the Voronoi cell $V_e^i$. Therefore,
  we can write the value of $u$, $v$ or $\sigma_{ij}$ at any point $(x,y)$ in $\widehat{V}_e^i$ as the linear combination of  basis functions as
 \begin{equation}\label{eq:basis2}
     f(x,y) = \sum_{j=1}^{n_c} f_{c_j} \mathcal{B}_{c_j}(x,y),
 \end{equation}
 where $n_c$ is the number of vertices of polygon $\widehat{V}_e^i$, $f$ is again the field, $f_{c_j}$ is the value of the field at the $j$-th vertex of $\widehat{V}_e^i$ (which is an edge location of $V_c$) and $\mathcal{B}_{c_j}$ is the basis function associated with the polygon vertex $c_j$.
 We also require $\{\mathcal{B}_{c_j}\}$ to have the same properties as in \eqref{properties}.
 Note that once again we are omitting the explicit dependence on $e$ in Eq. \eqref{eq:basis2} for ease of notation.
 \begin{remark}
 The choice of basis functions for the implementation will be either Wachspress \cite{dasgupta2003interpolants, turner2021mpas} or piecewise linear (PWL) \cite{bailey2008piecewise,turner2021mpas} basis functions, as these two options are those currently available in MPAS-Seaice for the B-grid. Both options guarantee the properties in \eqref{properties}.
 Moreover, with the present approximation, the fields are globally continuous over the computational domain because both $\{\mathcal{L}_{t_j}\}$ and $\{\mathcal{B}_{c_j}\}$ are linear at the edges of their respective domains of definition for either the Wachspress choice or the PWL.
  \end{remark}
We continue by considering $D_{u,e}^1$ and therefore the integral in \eqref{eq:D1}, which can be rewritten as
\begin{equation}\label{eq:D1bis}
 D_{u,e}^1 = -\sum\limits_{i=1}^2\int_{T_e^i} \dfrac{\partial}{\partial u_e}\Big(\sigma_{11} \Big[ \dfrac{\partial u} {\partial x} - v \, C_1(r) \tan(\lambda)\Big]\Big) dA   -\sum\limits_{i=1}^{2}\int_{\widehat{V}_e^i}\dfrac{\partial}{\partial u_e}\Big( \sigma_{11} \Big[ \dfrac{\partial u} {\partial x} - v \, C_1(r) \tan(\lambda)\Big] \Big)dA.
 \end{equation}
 Substituting Eq. \eqref{eq:basis1} and \eqref{eq:basis2} in Eq. \eqref{eq:D1bis} we get
\begin{equation}\label{eq:D1tris}
\begin{aligned}
 D_{u,e}^1 = &-\sum\limits_{i=1}^{2}\int_{T_e^i}  \dfrac{\partial}{\partial u_e}\Big(\sum_{j=1}^{n_t} {\sigma_{11}}_{t_j} \mathcal{L}_{t_j} \Big[ \sum_{k=1}^{n_t} u_{t_k} \dfrac{\partial \mathcal{L}_{t_k}} {\partial x} - \sum_{k=1}^{n_t} v_{t_k} \mathcal{L}_{t_k} C_1(r) \tan(\lambda)\Big]\Big) dA  \\ &-\sum\limits_{i=1}^{2}\int_{\widehat{V}_e^i} \dfrac{\partial}{\partial u_e}\Big(\sum_{j=1}^{n_c} {\sigma_{11}}_{c_j} \mathcal{B}_{c_j} \Big[ \sum_{k=1}^{n_c} u_{c_k} \dfrac{\partial \mathcal{B}_{c_k}} {\partial x} - \sum_{k=1}^{n_c} v_{ c_k} \mathcal{B}_{c_k} C_1(r)\tan(\lambda)\Big]\Big) dA.
 \end{aligned}
 \end{equation}
 Applying the derivative, the above expression simplifies to
\begin{equation}\label{eq:D1quadris}
 D_{u,e}^1 = -\sum\limits_{i=1}^{2}\int_{T_e^i}  \sum_{j=1}^{n_t} {\sigma_{11}}_{t_j} \mathcal{L}_{t_j} \dfrac{\partial \mathcal{L}_{\bar{e}}} {\partial x} dA   -\sum\limits_{i=1}^{2}\int_{\widehat{V}_e^i} \sum_{j=1}^{n_c} {\sigma_{11}}_{c_j} \mathcal{B}_{c_j} \dfrac{\partial \mathcal{B}_{\bar{e}}} {\partial x}  dA,
 \end{equation}
 where $\bar{e}$ is the local index that corresponds to the global index of $e$.
 We define the following matrices
   \begin{equation}\label{eq:matrices}
 \begin{aligned}
     (\mathbb{N}^x_{t_e^i})_{j,k} &= \int_{T_e^i}  \mathcal{L}_{t_j}  \dfrac{\partial \mathcal{L}_{t_k}}{\partial x}  dA , \qquad j,k=\{1,\ldots,n_t\}, \quad i=1,2,\\
         (\mathbb{N}_{v_e^i}^x)_{j,k} &= \int_{\widehat{V}_e^i}  \mathcal{B}_{c_j}  \dfrac{\partial \mathcal{B}_{c_k}}{\partial x}  dA , \qquad j,k=\{1,\ldots,n_c\}, \quad i=1,2,
         \end{aligned}
         \end{equation}
 where $n_t=3$ is the number of vertices of the triangle $T_e^i$ and $n_c$ is the number of vertices of $\widehat{V}_e^i$, e.g. $n_c=6$ for the orange hexagon in Fig. \ref{fig:2_bis}. 
         Then Eq. \eqref{eq:D1quadris} gives
         \begin{equation}\label{eq:D1final}
 D_{u,e}^1 = -\sum\limits_{i=1}^{2} \Big(\sum_{j=1}^{n_t} {\sigma_{11}}_{t_j}  (\mathbb{N}^x_{t_e^i})_{j,\bar{e}} +\sum_{j=1}^{n_c} {\sigma_{11}}_{c_j} (\mathbb{N}^x_{v_e^i})_{j,\bar{e}}\Big).
 \end{equation}
  For $D^1_{v,e}$, the computations are analogous until Eq. \eqref{eq:D1tris}, after which differentiation with respect to $v_e$ gives
  \begin{equation}\label{eq:D1v}
 D_{v,e}^1 = \sum\limits_{i=1}^{2}\int_{T_e^i}  \sum_{j=1}^{n_t} {\sigma_{11}}_{t_j} \mathcal{L}_{t_j} \mathcal{L}_{\bar{e}}\,C_1(r)\tan(\lambda)dA   +\sum\limits_{i=1}^{2}\int_{\widehat{V}_e^i} \sum_{j=1}^{n_c} {\sigma_{11}}_{c_j} \mathcal{B}_{c_j}  \mathcal{B}_{\bar{e}}\, C_1(r)\tan(\lambda) dA.
 \end{equation}
 To simplify the computations, we assume that  $\lambda$ varies slowly within $T_e^i$ or $\widehat{V}_e^i$, and that its value in these domains can be approximated by $\lambda_e$, the latitude at edge $e$. This approximation becomes less stringent as the computational cells size approaches zero. We introduce the following matrices
    \begin{equation}\label{eq:matrices2}
 \begin{aligned}
     (\mathbb{M}_{t_e^i})_{j,k} &= \int_{T_e^i}  \mathcal{L}_{t_j} \mathcal{L}_{t_k}  dA , \qquad j,k=\{1,\ldots,n_t\}, \quad i=1,2,\\
         (\mathbb{M}_{v_e^i})_{j,k} &= \int_{\widehat{V}_e^i}  \mathcal{B}_{c_j}  \mathcal{B}_{c_k} dA , \qquad j,k=\{1,\ldots,n_c\}, \quad i=1,2.
         \end{aligned}
         \end{equation}
 Then, Eq. \eqref{eq:D1v} becomes
   \begin{equation}\label{eq:D1v_final}
 D_{v,e}^1 = \sum\limits_{i=1}^{2}C_1(r)\tan(\lambda_e)\Big(  \sum_{j=1}^{n_t} {\sigma_{11}}_{t_j} (\mathbb{M}_{t_e^i})_{j,\bar{e}}   + \sum_{j=1}^{n_c} {\sigma_{11}}_{c_j} (\mathbb{M}_{v_e^i})_{j,\bar{e}}\Big).
 \end{equation}
The computations for $D^i_{u,e}$ and $D^i_{v,e}$ for $i=2,3$ are similar and are reported in Appendix A for completeness. Once those quantities are computed, the components of the divergence of the stress are then evaluated using Eq. \eqref{eq:integrated_eqs3} as follows
\begin{equation}\label{eq:final_div_stress}
\begin{aligned}
    F_{ue} = -\dfrac{1}{A_{P_e}} \sum\limits_{i=1}^{2}\Big[ &\Big(\sum_{j=1}^{n_t} {\sigma_{11}}_{t_j}  (\mathbb{N}^x_{t_e^i})_{j,\bar{e}} +\sum_{j=1}^{n_c} {\sigma_{11}}_{c_j} (\mathbb{N}^x_{v_e^i})_{j,\bar{e}}\Big)\\
     &+\Big( \sum_{j=1}^{n_t} {\sigma_{12}}_{t_j} \Big( (\mathbb{N}^y_{t_e^i})_{i,\bar{e}} + C_2(r)\tan(\lambda_{e})(\mathbb{M}_{t_e^i})_{j,\bar{e}} \Big) 
    +\sum_{j=1}^{n_c} {\sigma_{12}}_{c_j} \Big( (\mathbb{N}^y_{v_e^i})_{i,\bar{e}} + C_2(r)\tan(\lambda_{e})(\mathbb{M}_{v_e^i})_{i,\bar{e}} \Big) 
    \Big].
    \end{aligned}
\end{equation}
\begin{equation}\label{eq:final_div_stress2}
\begin{aligned}
    F_{ve} = -\dfrac{1}{A_{P_e}} \sum\limits_{i=1}^{2}\Big[- &C_1(r)\tan(\lambda_e)\Big(  \sum_{j=1}^{n_t} {\sigma_{11}}_{t_j} (\mathbb{M}_{t_e^i})_{j,\bar{e}}   + \sum_{j=1}^{n_c} {\sigma_{11}}_{c_j} (\mathbb{M}_{v_e^i})_{j,\bar{e}}\Big)\\
     &+\Big(\sum_{j=1}^{n_t}  {\sigma_{12}}_{t_j}  (\mathbb{N}^x_{t_e^i})_{j,\bar{e}}+     \sum_{j=1}^{n_c}  {\sigma_{12}}_{c_j}  (\mathbb{N}^x_{v_e^i})_{j,\bar{e}} \Big) \\
     &+\Big( \sum_{j=1}^{n_t}  {\sigma_{22}}_{t_j}\Big(  (\mathbb{N}^y_{t_e^i})_{j,\bar{e}}+C_3(r)\tan(\lambda_{e})(\mathbb{M}_{t_e^i})_{j,\bar{e}}\Big)
 +\sum_{j=1}^{n_c}  {\sigma_{22}}_{c_j}  \Big((\mathbb{N}^y_{v_e^i})_{j,\bar{e}}+C_3(r)\tan(\lambda_{e})(\mathbb{M}_{v_e^i})_{j,\bar{e}}\Big)
    \Big].
    \end{aligned}
\end{equation}
For edges on the boundary, the velocity components are set to zero, hence the stress values ${\sigma_{11}}_{t_j}$ are zero. This means that the computation of the matrices in \eqref{eq:matrices} and \eqref{eq:matrices2} is allowed to be inexact for those triangles associated with boundary vertices whose vertices (which are primal edge points) are not all part of the mesh (see Figure \ref{fig:2_bis}). Such inexact value is then multiplied by zero in Eq. \eqref{eq:D1final} and Eq. \eqref{eq:D1v_final}, and so it does not contribute to the divergence of the stress. Note that the computation of the matrices for the blue triangles is correct since all their vertices are available on the mesh. Moreover, when continents are considered, the actual boundary of the domain for the computation of the divergence of the stress is given by the edges of the blue-grey triangles in Figure \ref{fig:2_bis}.
 }
 \begin{remark}
 We observe that the proposed approach applied to a structured quadrilateral grid is equivalent to discretizing the momentum equation on a rotated grid, to relocate the velocity components from the vertices to the edges, as it is clear from Figure \ref{fig:2} (left).
 \end{remark}
 
 \subsubsection{Consistent approach}\label{sec:alternative}
\Giacomo{As observed in \cite{turner2021mpas} where the unstructured B-grid formulation was presented, it is possible to define an alternative formulation to that presented in section \ref{sec:standard} by using the basis expansions for the right hand side integral in Eq. \eqref{eq:integrated_eqs}, as well as for the integral on the left hand side. This is actually the consistent approach from a mathematical point of view, and while both approaches are convergent for the B-grid formulation, it has been shown in \cite{turner2021mpas} that for the B-grid the consistent approach produces smaller errors than the standard one on a unit sphere for the divergence of the stress operator. On the other hand, the standard approach is convergent on planar test cases for the CD-grid, but the consistent approach is the only one capable of converging on the unit sphere.
For brevity, considering Eq. \eqref{eq:integrated_eqs}, we focus only on the left hand side integral involving $\partial/ \partial u$ because the procedure is analogous for the one involving $\partial/ \partial v$. We have
\begin{equation}
\begin{aligned}
         \int_{\mathcal{V}_e} \dfrac{\partial}{\partial u}\Big(u\,F_u + v\,F_v\Big) \,dA = \sum_{i=1}^2\Big[ \int_{T_e^i} \dfrac{\partial}{\partial u}\Big(u\,F_u + v\,F_v\Big) \,dA +  \int_{\widehat{V}_e^i} \dfrac{\partial}{\partial u}\Big(u\,F_u + v\,F_v\Big) \,dA\Big].
         \end{aligned}
\end{equation}
Introducing the basis expansions, the above equation becomes
\begin{equation} 
\begin{aligned}
               \int_{\mathcal{V}_e} \dfrac{\partial}{\partial u}\Big(u\,F_u + v\,F_v\Big) \,dA =   \sum\limits_{i=1}^{2}&\Big[\int_{T_e^i}   \dfrac{\partial}{\partial u_e}\Big(\sum_{j=1}^{n_t} u_{{t_j}} \mathcal{L}_{t_j}\Big)\Big(\sum_{k=1}^{n_t} F_{u_{t_k}} \mathcal{L}_{t_k}\Big) +  \dfrac{\partial}{\partial u_e}\Big(\sum_{z=1}^{n_t} v_{{t_z}} \mathcal{L}_{t_z}\Big)\Big(\sum_{l=1}^{n_t} F_{v_{t_l}} \mathcal{L}_{t_l}\Big) dA,\\
             &\int_{\widehat{V}_e^i}   \dfrac{\partial}{\partial u_e}\Big(\sum_{j=1}^{n_c} u_{{c_j}} \mathcal{B}_{c_j}\Big)\Big(\sum_{k=1}^{n_c} F_{u_{c_k}} \mathcal{B}_{c_k}\Big) +  \dfrac{\partial}{\partial u_e}\Big(\sum_{z=1}^{n_c} v_{{c_z}} \mathcal{B}_{c_z}\Big)\Big(\sum_{l=1}^{n_c} F_{v_{c_l}} \mathcal{B}_{c_l}\Big) dA\Big].
\end{aligned}
\end{equation}
After applying the derivatives we obtain
\begin{equation} 
\begin{aligned}
               \int_{\mathcal{V}_e} \dfrac{\partial}{\partial u}\Big(u\,F_u + v\,F_v\Big) \,dA =    \sum\limits_{i=1}^{2}&\Big[\int_{T_e^i}   \sum_{k=1}^{n_t} F_{u_{t_k}} \mathcal{L}_{t_k}\mathcal{L}_{t_{\bar{e}}} \, dA + \int_{\widehat{V}_e^i}\sum_{k=1}^{n_c} F_{u_{c_k}} \mathcal{B}_{c_k}\mathcal{B}_{c_{\bar{e}}} \,dA\Big],
\end{aligned}
\end{equation}
  with $\bar{e}$ being again the local index corresponding to $e$. 
We now make the approximation that $F_u$ varies slowly spatially within $\mathcal{V}_e$ and that its value in this set can be approximated with its value at $e$, i.e. $F_{ue}$. Hence the above equation simplifies to\begin{equation} 
\begin{aligned}
               \int_{\mathcal{V}_e} \dfrac{\partial}{\partial u}\Big(u\,F_u + v\,F_v\Big) \,dA= F_{ue} \, \sum\limits_{i=1}^{2}&\Big[\int_{T_e^i}   \sum_{k=1}^{n_t}  \mathcal{L}_{t_k}\mathcal{L}_{t_{\bar{e}}} \, dA+ \int_{\widehat{V}_e^i}\sum_{j=1}^{n_c}  \mathcal{B}_{c_j}\mathcal{B}_{c_{\bar{e}}} \,dA\Big],
\end{aligned}
\end{equation}
Using the first property in Eq. \eqref{properties}, this leads to Eq. \eqref{eq:integrated_eqs3}, with $A_{P_e}$ given by 
  \begin{align}\label{eq:altDen}
      A_{P_e} = \sum\limits_{i=1}^{2}&\Big[\int_{T_e^i}   \mathcal{L}_{t_{\bar{e}}} \, dA+ \int_{\widehat{V}_e^i}\mathcal{B}_{c_{\bar{e}}} \,dA\Big],
  \end{align}
  instead of being the area of the diamond-shaped polygon associated with edge $e$}.
  
  \subsection{Differences with existing methods}
  \Giacomo{The approach proposed in this work differs from the CD-grid method developed by Mehlmann and Korn (MK) \cite{mehlmann2021sea} for several reasons.
  First, in \cite{mehlmann2021sea} a finite element (FE) method is used, whereas here it is not. In our approach there is no linear system that is solved and the velocity is not really expressed as the linear combination of basis functions in a FE sense, but rather the expansion is only used to compute the integrals for the divergence of the internal stress at the discretization points, and does not play a role in the overall solution of the dynamics equation.
There are also differences between the basis functions used here for the expansion of the velocity and those employed by MK.  Namely, let us consider the edge location marked with two red triangles in Figure \ref{fig:comparison}: 
\begin{figure}[h]
   \centering
   \includegraphics[scale=0.7]{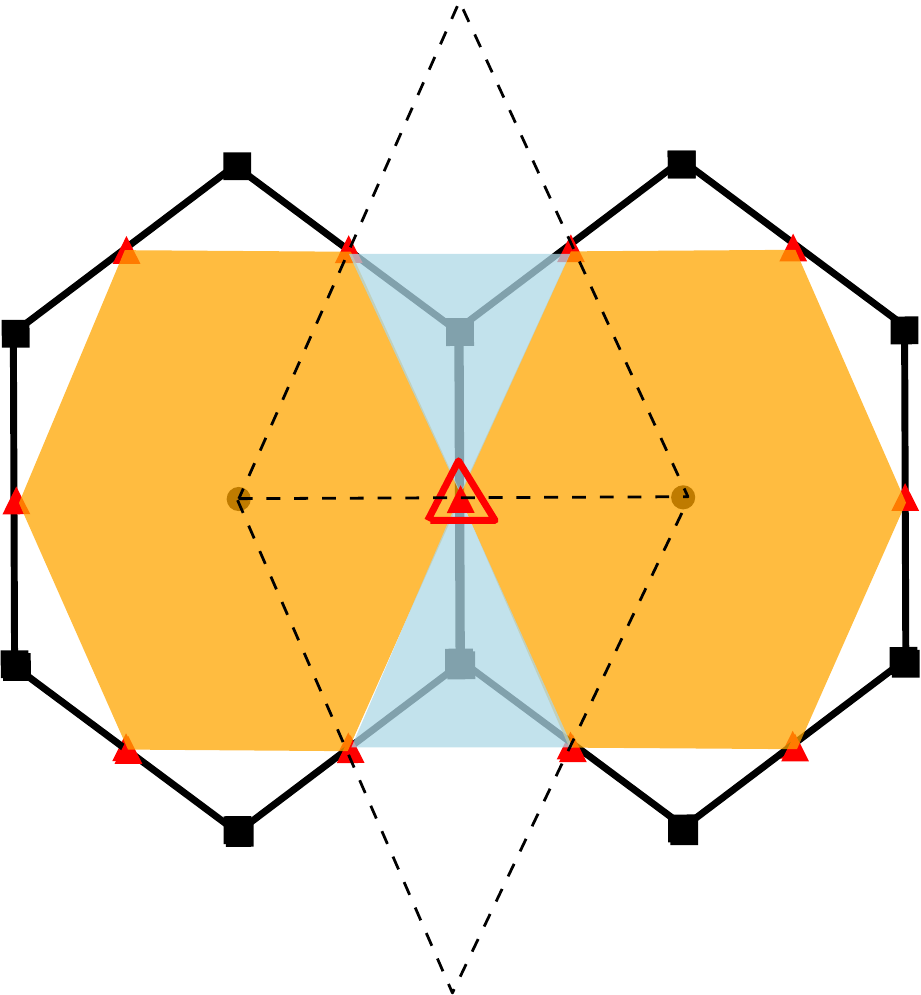}
\caption{Support of the basis functions with the present method (colored shapes), and with the method in \cite{mehlmann2021sea} (dashed triangles). The dashed triangles are also examples of cells of the dual mesh.}
   \label{fig:comparison}
\end{figure}
 with the approach by MK, the (global) basis function associated with this edge has support on the two dashed triangles (which are part of the dual mesh) and has a value of one on the dashed line segment that goes through it. In the present framework, considering the consistent approach in Section \ref{sec:alternative}, the support is given by the union of the two orange polygons and the two blue-gray triangles. The global basis function would be continuous across the edges of the orange and blue-gray polygons (with either Wachspress or PWL, since they are both linear at the edges) but it would not have a value of one of the dashed line segment, rather only on the edge location (the point marked with two red triangles).
 Moreover, the basis functions from MK have range $[-1,1]$, whereas in our approach Wachspress and PWL only attain non-negative values. 
 It is true though that the MK basis function restricted to the blue-grey triangles do coincide with the basis function we use on such blue-grey triangles.
 Another difference is that MK expand  the velocity with basis functions using a FE approach, but do not expand the stress or its divergence. This is consistent with the FE framework chosen by MK. Finally, the method in in \cite{mehlmann2021sea} needs a stabilization term whereas ours does not.}
  
 \section{Numerical Results}\label{sec:numRes}
 To investigate the properties of our formulation, we present results on the plane and sphere focusing on the accuracy and convergence of the proposed discretization using analytical solutions as references.

 \subsection{Spatial discretization test}\label{sec:spat_disc}
We begin with a theoretical analysis to obtain sufficient conditions under which the proposed scheme is expected to be at least a second-order approximation of the continuous divergence operator.
Placing ourselves in a general setting, let us consider $\Omega \subset \mathbf{R}^2$ to be a closed and bounded set, and let $\bm{\sigma}$ be a stress tensor defined on $\Omega$ given by
 \begin{equation}
 \bm{\sigma}=
 \begin{bmatrix}
\sigma_{11} & \sigma_{12} \\
\sigma_{21} & \sigma_{22} 
\end{bmatrix},
\end{equation}
with $\sigma_{12} = \sigma_{21}$.
Note that as before, $\Omega$ denotes the computational domain. We also introduce $\Omega_e \subset \Omega$ to be the set of all edge points of the mesh, i.e. if $e$ is an edge of the mesh, then $(x_e,y_e) \in \Omega_e$.
The divergence of the stress is a $1\times2$ vector defined as
 \begin{equation}
 \nabla \cdot \bm{\sigma}=
 \begin{bmatrix}
\dfrac{\partial \sigma_{11}}{\partial x}  \,\, + \,\,  \dfrac{\partial \sigma_{21}}{\partial y} \\
\dfrac{\partial \sigma_{12}}{\partial x} \,\,  + \,\,  \dfrac{\partial \sigma_{22}}{\partial y}   
\end{bmatrix}^T.
\end{equation}
Above, we have defined $\bm{F}=(F_{u},F_{v}) := \nabla \cdot \bm{\sigma}$, hence
\begin{equation}
    F_{u} = \dfrac{\partial \sigma_{11}}{\partial x}  \,\, + \,\,  \dfrac{\partial \sigma_{21}}{\partial y}, \qquad F_{v} = \dfrac{\partial \sigma_{12}}{\partial x} \,\,  + \,\,  \dfrac{\partial \sigma_{22}}{\partial y}.
\end{equation}
Evaluated at an edge location $(x_e,y_e) \in \Omega_e$, the above functions give
\begin{equation}\label{eq:eq1ToCompare}
    F_{u}(x_e,y_e) = \dfrac{\partial \sigma_{11}}{\partial x}(x_e,y_e)  \,\, + \,\,  \dfrac{\partial \sigma_{21}}{\partial y}(x_e,y_e), \qquad F_{v}(x_e,y_e) = \dfrac{\partial \sigma_{12}}{\partial x}(x_e,y_e) \,\,  + \,\,  \dfrac{\partial \sigma_{22}}{\partial y}(x_e,y_e).
\end{equation}
\Giacomo{For the same point $(x_e,y_e) \in \Omega_e$, for a planar case with no metric terms, Eq.\eqref{eq:final_div_stress} gives
\begin{equation}
\begin{aligned}\label{eq:discrete_Fue}
    F_{ue} = -\dfrac{1}{A_{P_e}} \sum\limits_{i=1}^{2}\Big( \sum_{j=1}^{n_t} {\sigma_{11}}_{t_j}  (\mathbb{N}^x_{t_e^i})_{j,\bar{e}} +\sum_{j=1}^{n_c} {\sigma_{11}}_{c_j} (\mathbb{N}^x_{v_e^i})_{j,\bar{e}}+ \sum_{j=1}^{n_t} {\sigma_{12}}_{t_j}  (\mathbb{N}^y_{t_e^i})_{i,\bar{e}} 
    +\sum_{j=1}^{n_c} {\sigma_{12}}_{c_j}  (\mathbb{N}^y_{v_e^i})_{i,\bar{e}} 
    \Big).
    \end{aligned}
\end{equation}
Recall from above that $\bar{e}$ is the local index associated to the edge $e$.
It follows by comparing \eqref{eq:eq1ToCompare} and \eqref{eq:discrete_Fue} that with our spatial discretization of the divergence of the stress, we are approximating the gradient operator at the edge points $(x_e,y_e) \in \Omega_e$ in the following way:
\begin{equation}\label{eq:approx1}
\begin{aligned}
    \dfrac{\partial \sigma_{11}}{\partial x}(x_e,y_e)  &\approx -\dfrac{1}{A_{P_e}} \sum\limits_{i=1}^{2}\Big( \sum_{j=1}^{n_t} {\sigma_{11}}_{t_j}  (\mathbb{N}^x_{t_e^i})_{j,\bar{e}} +\sum_{j=1}^{n_c} {\sigma_{11}}_{c_j} (\mathbb{N}^x_{v_e^i})_{j,\bar{e}} \Big),\\
        \dfrac{\partial \sigma_{21}}{\partial y}(x_e,y_e) &\approx -\dfrac{1}{A_{P_e}} \sum\limits_{i=1}^{2}\Big( \sum_{j=1}^{n_t} {\sigma_{12}}_{t_j}  (\mathbb{N}^y_{t_e^i})_{j,\bar{e}} +\sum_{j=1}^{n_c} {\sigma_{12}}_{c_j} (\mathbb{N}^y_{v_e^i})_{j,\bar{e}} \Big).
    \end{aligned}
\end{equation}.
Similarly for $F_{ve}$ in Eq.\eqref{eq:final_div_stress2}, still considering a planar case with no metric terms, we have
\begin{equation}
\begin{aligned}
    F_{ve} = -\dfrac{1}{A_{P_e}} \sum\limits_{i=1}^{2}\Big[\sum_{j=1}^{n_t}  {\sigma_{12}}_{t_j}  (\mathbb{N}^x_{t_e^i})_{j,\bar{e}}+     \sum_{j=1}^{n_c}  {\sigma_{12}}_{c_j}  (\mathbb{N}^x_{v_e^i})_{j,\bar{e}}  
   +\sum_{j=1}^{n_t}  {\sigma_{22}}_{t_j}  (\mathbb{N}^y_{t_e^i})_{j,\bar{e}}
 +\sum_{j=1}^{n_c}  {\sigma_{22}}_{c_j}  (\mathbb{N}^y_{v_e^i})_{j,\bar{e}}
    \Big].
     \end{aligned}
\end{equation}}
\Giacomo{This implies that
\begin{equation}\label{eq:approx2}
\begin{aligned}
        \dfrac{\partial \sigma_{12}}{\partial x}(x_e,y_e) &\approx  -\dfrac{1}{A_{P_e}} \sum\limits_{i=1}^{2}\Big[\sum_{j=1}^{n_t}  {\sigma_{12}}_{t_j}  (\mathbb{N}^x_{t_e^i})_{j,\bar{e}}+     \sum_{j=1}^{n_c}  {\sigma_{12}}_{c_j}  (\mathbb{N}^x_{v_e^i})_{j,\bar{e}}\Big],\\
    \dfrac{\partial \sigma_{22}}{\partial y}(x_e,y_e)  &\approx -\dfrac{1}{A_{P_e}} \sum\limits_{i=1}^{2}\Big[\sum_{j=1}^{n_t}  {\sigma_{22}}_{t_j}  (\mathbb{N}^y_{t_e^i})_{j,\bar{e}}
 +\sum_{j=1}^{n_c}  {\sigma_{22}}_{c_j}  (\mathbb{N}^y_{v_e^i})_{j,\bar{e}}\Big],
    \end{aligned}
\end{equation}
which leads to the same conclusion reached after Eq. \eqref{eq:approx1}, i.e. that our spatial discretization approximates the gradient of a function with a linear functional $\mathcal{F} : C^{2}(\Omega) \rightarrow \mathcal{B}(\Omega_e)$ such that for any $f \in C^{2}(\Omega)$ we have 
\begin{equation}\label{eq:gradApprox}
\nabla f(x_e,y_e) \approx \mathcal{F}(f)(x_e,y_e) :=
 \begin{bmatrix}
-\dfrac{1}{A_{P_e}} \sum_{i=1}^2\Big(\sum_{j=1}^{n_t}  f_{t_j} (\mathbb{N}^x_{t_e^i})_{j,\bar{e}} 
    \,\,+\sum_{j=1}^{n_c}  f_{c_j} (\mathbb{N}^x_{v_e^i})_{j,\bar{e}}\Big)    \vspace{0.5cm}\\
-\dfrac{1}{A_{P_e}} \sum_{i=1}^2\Big(\sum_{j=1}^{n_t}  f_{t_j} (\mathbb{N}^y_{t_e^i})_{j,\bar{e}} 
    \,\,+\sum_{j=1}^{n_c}  f_{c_j} (\mathbb{N}^y_{v_e^i})_{j,\bar{e}}\Big)     
\end{bmatrix},
\end{equation}}
where $C^{2}(\Omega)$ is the space of twice differentiable functions with continuous derivatives on $\Omega$, and $\mathcal{B}(\Omega_e)$ denotes the space of bounded functions on $\Omega_e$. 
\Giacomo{Let $\mathbf{y}=(y_1,y_2) \in \Omega_e$ and $\mathbf{x}=(x_1,x_2) \in \Omega_e$ and define 
\begin{align}
g_1(\mathbf{x}) := 1,& \qquad g_2(\mathbf{x}): = (x_1-y_1), \qquad g_3(\mathbf{x}): = (x_2-y_2), \qquad
g_4(\mathbf{x}) := (x_1-y_1)^2,\\
\qquad g_5(\mathbf{x}) := &(x_2-y_2)(x_1-y_1), \qquad g_6(\mathbf{x})=g_5(\mathbf{x}), \qquad g_7(\mathbf{x}) = (x_2-y_2)^2.
\end{align}
Then, using a Taylor expansion argument whose details are reported in Appendix C, we obtain the following sufficient conditions  for $\mathcal{F}$ to be a at least a second-order approximation of $\nabla f$
\begin{equation}\label{eq:conditions}
    \begin{cases}
    \mathcal{F}(g_i)((x_e,y_e)) = [0,0]^T, i\neq 2, i \neq3, \\
    \mathcal{F}(g_2)((x_e,y_e)) = [1,0]^T, \\
    \mathcal{F}(g_3)((x_e,y_e)) = [0,1]^T.
    \end{cases},
\end{equation}
for any point $(x_e,y_e) \in \Omega_e$.}
Note that if $\mathcal{F}$ was indeed equal to $\nabla f$ then the above conditions would be satisfied.
With respect to the analysis just concluded, we have numerically estimated the values of $\mathcal{F}(g_i)$ for $i=\{1,2,3,4,5,7\}$, considering $\mathcal{F}$ to be the operator obtained with the CD-grid approach proposed in this paper and also the one obtained with the B-grid approach form \cite{turner2021mpas}.
We use a planar mesh with regular hexagonal cells and one with square cells.
For the B-grid approach, we select a vertex of the mesh that is not on the boundary of the domain, and also not surrounded by boundary cells. For the CD-grid case, we consider the edges that have such a vertex in common.
With this setup, for the CD-grid case there will be three edges for the hexagonal mesh, each of which will be oriented differently, and four edges for the quadrilateral mesh, with pairs of edges oriented in the same way.
%The meshes in consideration and relative vertices and edges are shown in Figure \ref{fig:op_meshes1} for the mesh with square cells and Figure \ref{fig:op_meshes2} for the mesh with hexagonal cells.
To avoid numerical error, instead of computing the first component of $\mathcal{F}(g_2)$, we compute the difference
$$\Big|A_{P_e}- \Big(- \sum_{i=1}^2\Big(\sum_{j=1}^{n_t}  {g_2}_{t_j} (\mathbb{N}^x_{t_e^i})_{j,\bar{e}} 
    \,\,+\sum_{j=1}^{n_c}  {g_2}_{c_j} (\mathbb{N}^x_{v_e^i})_{j,\bar{e}}\Big)\Big| , $$
    hence if such a  difference is zero, then the first component of $\mathcal{F}(g_2)$ is one. Let us denote with $\widetilde{\mathcal{ F}}(g_2)$ the vector $\mathcal{F}(g_2)$ whose first entry has been modified as explained. We adopt the same strategy for the second entry of $\mathcal{F}(g_3)$, and define $\widetilde{\mathcal{ F}}(g_3)$ in a similar way as $\widetilde{\mathcal{ F}}(g_2)$.
    Hence, if $\widetilde{\mathcal{ F}}(g_i)=\mathbf{0}$ for $i=2,3$ and $\mathcal{F}(g_j)=\mathbf{0}$ for $j=1,4,5,7$, then according to the conditions in \eqref{eq:conditions}, we can expect the methods to be at least second-order accurate. Obviously the equality to the zero vector is intended in the machine precision sense.
    For this test (and all the planar tests) the area $A_{P_e}$ will be the area of the diamond-like shapes in Figure \ref{fig:2_bis} and Figure \ref{fig:2} (right).  We consider both Wachspress and PWL basis functions.
    For the mesh with square cells, the B-grid returned zero vectors for both Wachspress and PWL basis functions, as did the CD-grid, for all four edges considered and both types of basis functions. Hence, on the quadrilateral mesh with square cells, both methods are expected to show second-order convergence for the divergence of the stress operator.
    For the hexagonal mesh, the CD-grid returned zero vectors for all three edges and both choices of basis functions, whereas the B-grid did so only for the PWL basis functions. In fact, with the Wachspress choice we had 
    \begin{equation}
        \mathcal{F}(g_4) = [0,\gamma_1]^T, \quad  \mathcal{F}(g_5) = [\gamma_2,0]^T, \quad   \mathcal{F}(g_7) = [0,\gamma_3]^T,
    \end{equation}
    with $\gamma_1,\gamma_2$ and $\gamma_3$ being non zero numbers.
    Hence, the choice of Wachspress basis functions is not expected to be second-order with the B-grid approach but only at least first order. This was already observed in \cite{turner2021mpas}.
    %%%%%%%%%%%%
%\begin{figure}[!t]
%   \centering
%   \includegraphics[scale=0.4]{figures/quad_mesh}
%\caption{Quadrilateral mesh used for the spatial discretization test. The vertex used for the B-grid is number 5071. For the C-grid the edges that join at vertex 5071 are considered.}
%   \label{fig:op_meshes1}
%\end{figure}
%\begin{figure}[!t]
%   \centering
%   \includegraphics[scale=0.48]{figures/hex_mesh}
%\caption{Hexagonal mesh used for the spatial discretization test. The vertex used for the B-grid is number 1123. For the C-grid, the edges that join at vertex 1123 are considered.}
%   \label{fig:op_meshes2}
%\end{figure}
%%%%%%%%%%%%
    
    \subsection{Convergence rate test}
    We continue with two tests to assess the accuracy of the proposed discretization in approximating the divergence of the internal stress. Namely, we first consider a unit square domain discretized with the same planar meshes used in the previous section (although with hexagonal cells the domain is not exactly a unit square), and then move to a unit sphere domain  discretized with a Voronoi tessellation. In all cases, we consider the values of the internal stress to be prescribed (i.e. given as input) at the edges or vertices, and obtained analytically through the simplest constitutive relation,  $\bm{\sigma}_{ij} = {\dot {\bm{\epsilon}}}_{ij}$, i.e. we assume the strain rate and the stress to be equal.
     
    \subsubsection{Convergence rate test on a planar mesh}
\begin{figure}[!t]
   \centering
      \includegraphics[scale=0.28]{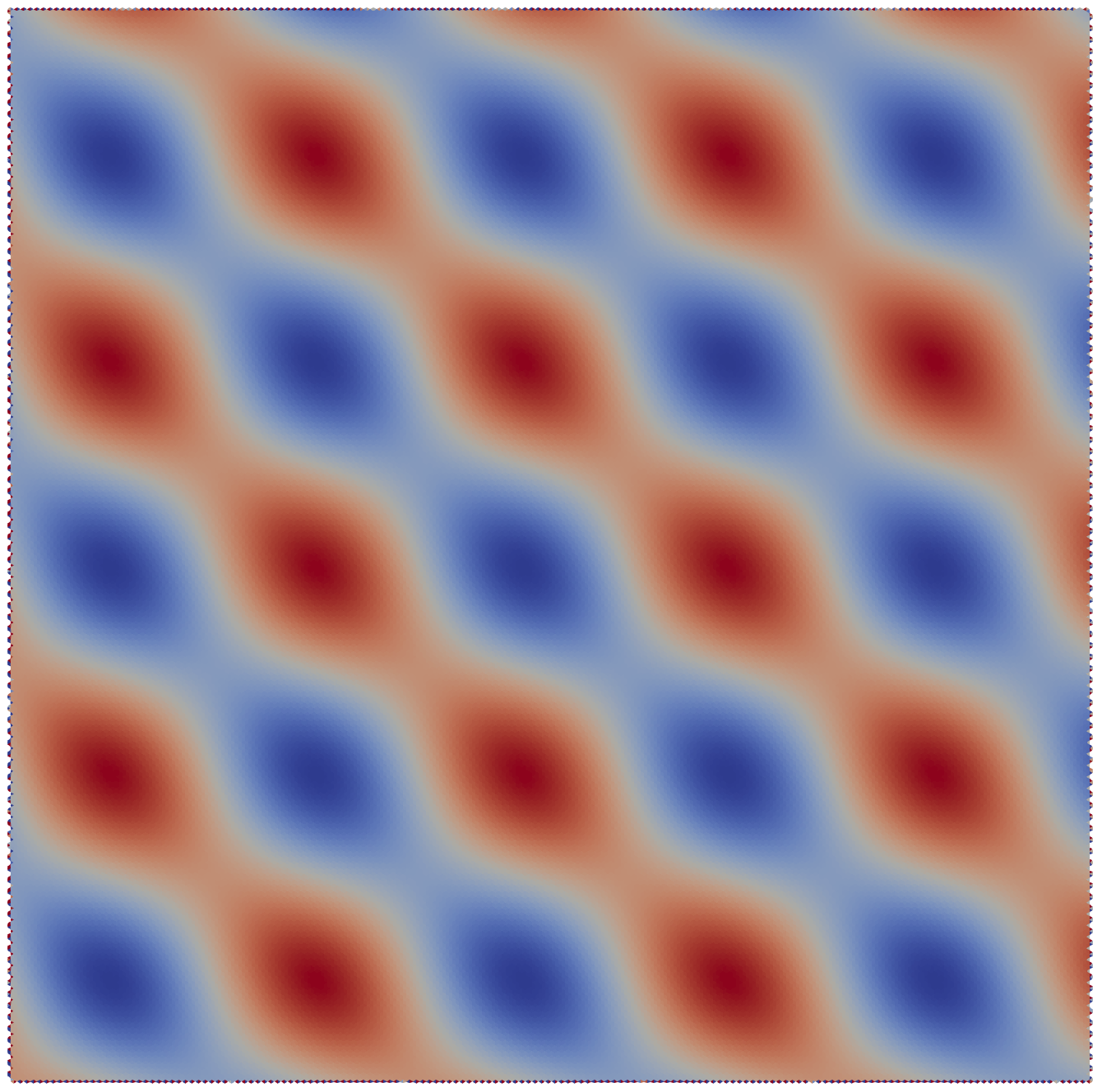}
   \includegraphics[scale=0.44]{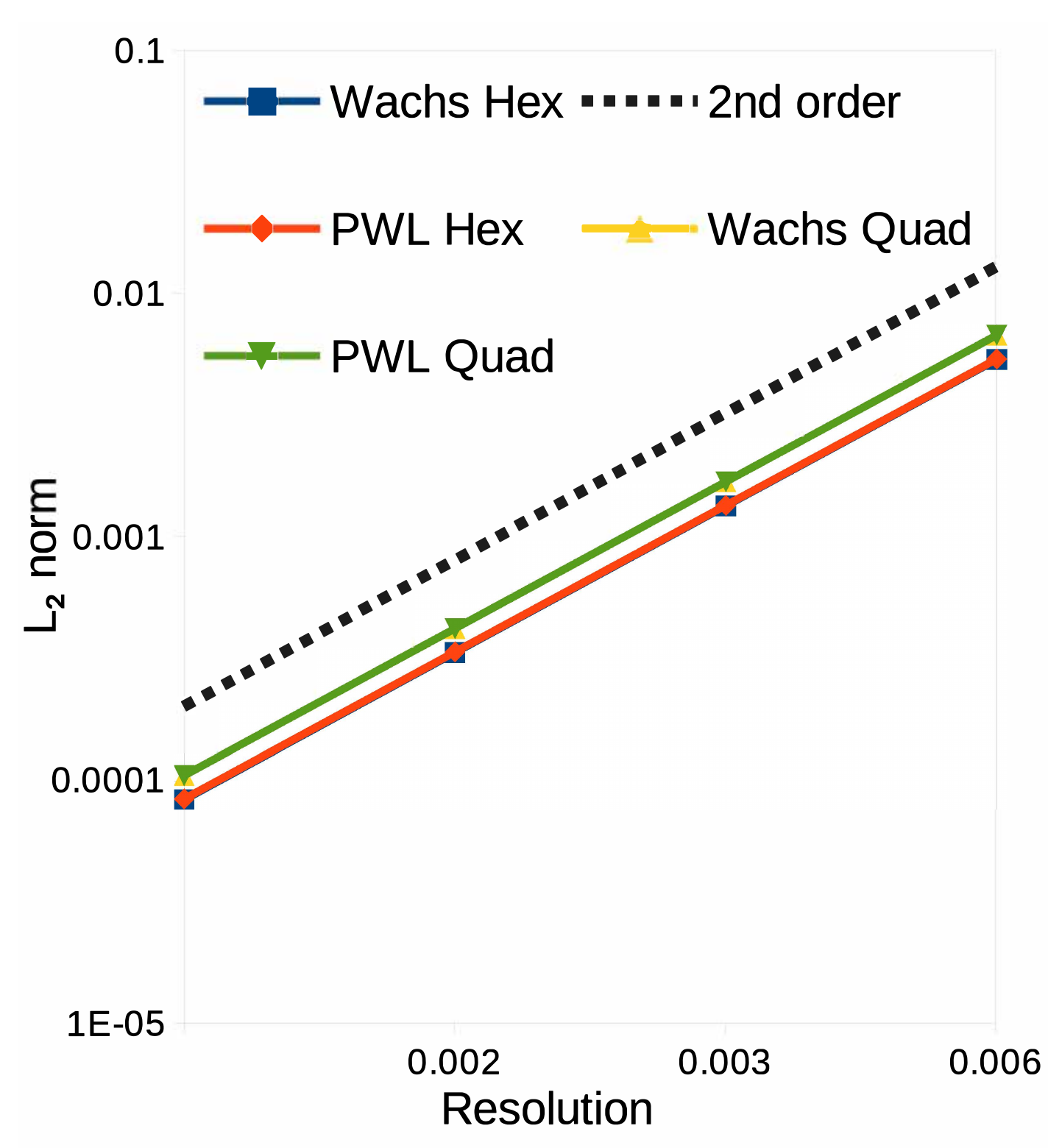}
\caption{Left: qualitative plot of the northward component of the divergence of the stress obtained with the CD-grid approach on the mesh with hexagonal cells. Right: convergence rate of the eastward component of the divergence of the stress using the CD-grid approach. Wachs: Wachspress basis. PWL: piecewise linear basis.}
   \label{fig:normUCGrid}
\end{figure}

\begin{figure}[!t]
   \centering
      \includegraphics[scale=0.5]{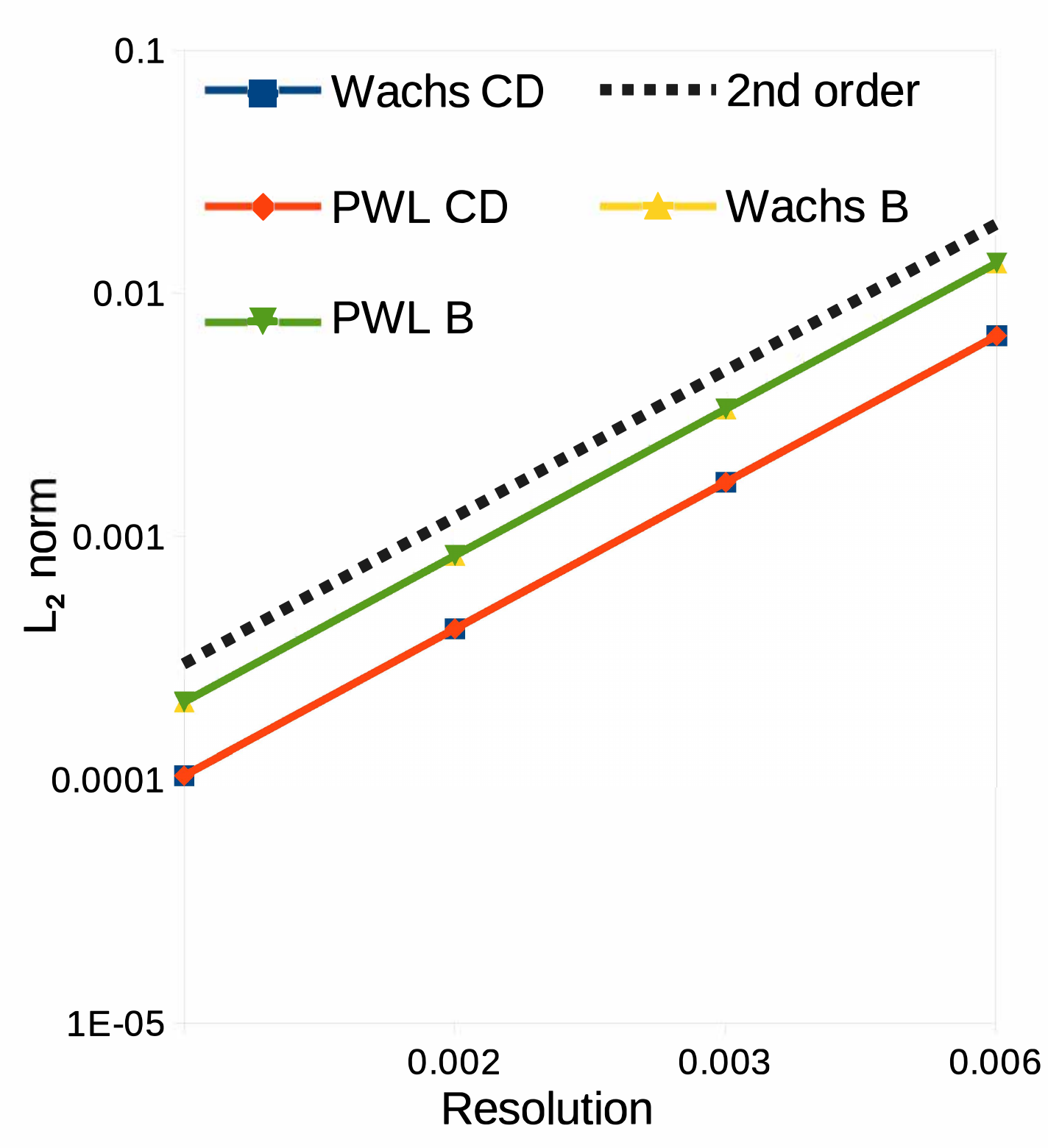}
   \includegraphics[scale=0.5]{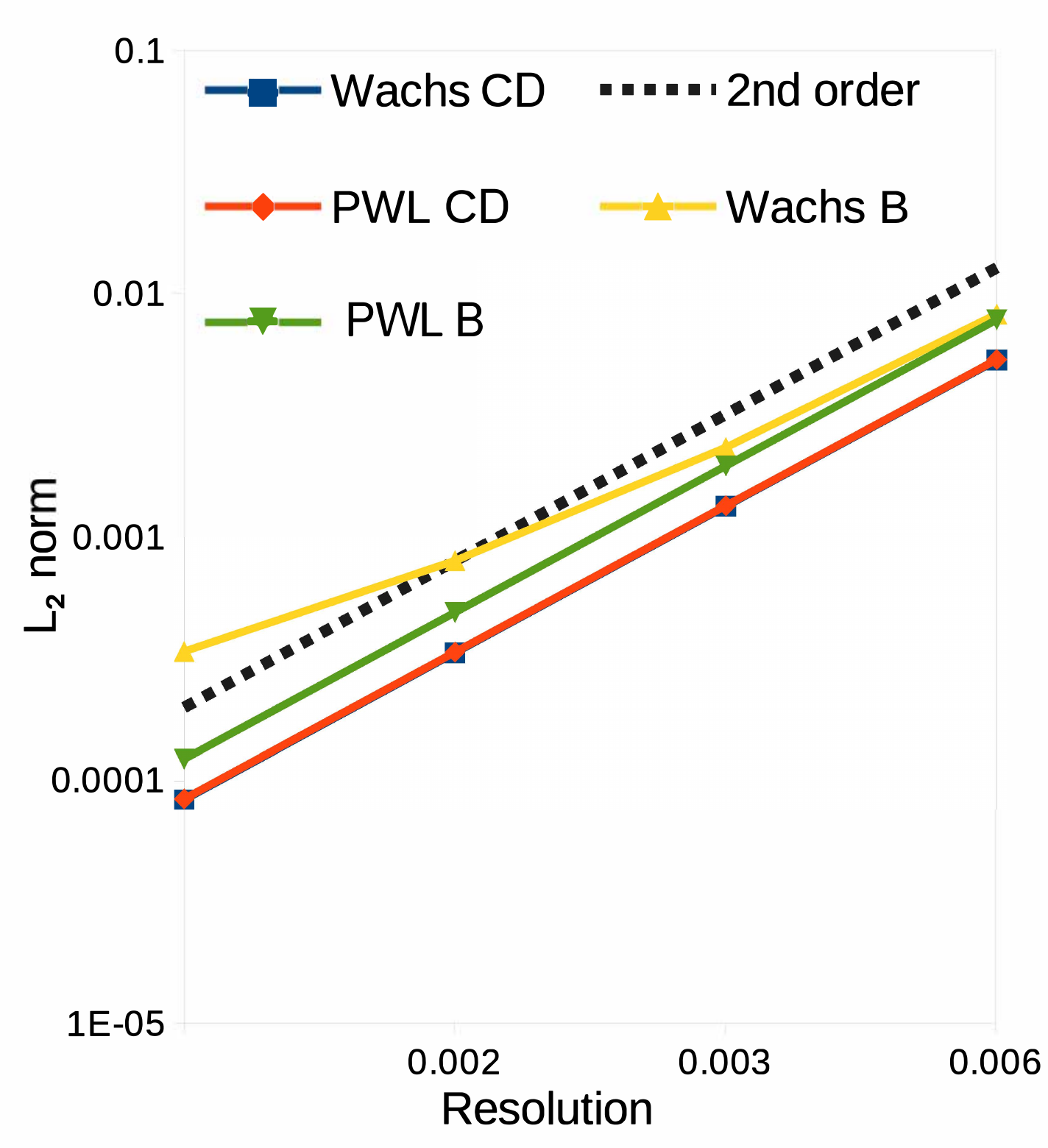}
\caption{Comparison of the convergence rate of the eastward component of the divergence of the stress using the CD-grid and the B-grid approach. Left: planar mesh with square cells. Right: planar mesh with regular hexagonal cells. Wachs: Wachspress basis. PWL: piecewise linear basis.}
 \label{normUComp}
\end{figure}

    For the planar mesh test case, the strain is obtained from derivatives of an analytical velocity field $\bm{u} = (u,v)$ given by
    \begin{equation}
     u(x,y) = \sin(5.12 \pi x)\sin(5.12 \pi y), \qquad v(x,y)=u(x,y).
     \end{equation}
     The strain rate (and hence the stress) is given by
     \begin{equation}
     \dot{\epsilon}_{11} = \dfrac{\partial u}{\partial x}, \quad \dot{\epsilon}_{22} = \dfrac{\partial u}{\partial y}, \quad
     \dot{\epsilon}_{12} = \frac{1}{2}\Big(\dfrac{\partial u}{\partial x} + \dfrac{\partial u}{\partial y}\Big). 
      \end{equation}
     Therefore, the analytical field we use to compute errors is
     \begin{equation}
         \bm{F} = \Big[\dfrac{\partial \dot \epsilon_{11}}{\partial x} + \dfrac{\partial \dot\epsilon_{12}}{\partial y}, \,\,\dfrac{\partial \dot\epsilon_{12}}{\partial x} + \dfrac{\partial \dot\epsilon_{22}}{\partial y} \Big].
     \end{equation}
     We are going to compare the B-grid and the CD-grid formulations using the following discrete relative $L_2$ norm
     \begin{equation}\label{eq:L2norm}
     \dfrac{\sqrt{\sum A_i (F_i - F^{\dag}_i)^2}}{\sqrt{\sum A_i F_i^2}}
     \end{equation}
where $F$ denotes any component of $\bm{F}$ and $F^{\dag}$ any component of the numerical  approximation ${\bm{F}^{\dag}}$ of $\bm{F}$ computed either with the B-grid or the CD-grid method. Note that  $\bm{F}^{\dag}$ is only available either at the vertices or at the edges, whereas $\bm{F}$ being an analytic field can be computed at any spatial location $(x,y)$. In Eq. \eqref{eq:L2norm}, the summation is taken over the vertices for the B-grid and over the edges for the CD-grid. For the planar case, $A_i$ will be the area of the dual triangle centered at the $i$-th vertex for the B-grid or the area of the diamond-shaped figure centered at the $i$
-th edge for the CD-grid.
Results for the CD-grid approach for both the mesh with square cells and the one with hexagonal cells are shown in Figure \ref{fig:normUCGrid} (right), considering the eastward component of the divergence of the stress (results are analogous for the northward component, hence they're not shown). 
 %%%%%%%%%%%%
%%%%%%%%%%%%
We observe that, as expected from the test in Section  \ref{sec:spat_disc}, the method has second-order convergence on both types of meshes and with both types of basis functions, i.e. Wachspress and PWL. Moreover, the choice of basis function does not affect the quality of the approximation, as the associated curves lie on top of each other for a given choice of mesh cells.
In Figure \ref{fig:normUCGrid} (left) we also display the qualitative behavior of the numerical solution obtained on the mesh with hexagonal cells for the CD-grid.
Next, we compare the B-grid and CD-grid approaches on the mesh with square cells and on the one with hexagonal cells, see Figure \ref{normUComp}.
We only show the behavior of the eastward component of the divergence of the stress, because the northward showed an analogous behavior.
 %%%%%%%%%%%%
%%%%%%%%%%%%
Once again, as expected by the analysis in Section \ref{sec:spat_disc}, the CD-grid approach and the B-grid approach are both second-order on the mesh with square cells, with the CD-grid showing lower errors than the B-grid.
Both methods are insensitive to the choice of basis functions in this case.
For the mesh with hexagonal cells, the CD-grid approach has again lower errors compared to the B-grid, for which the case of Wachspress basis functions becomes first order as the resolution of the mesh is increased. This behavior for the B-grid was expected from the analysis in the previous section and already reported in \cite{turner2021mpas}.

     \subsubsection{Convergence rate test on a spherical mesh}
We continue our analysis considering a spherical Voronoi mesh on a unit sphere. An investigation of the mesh quality revealed that most cells are hexagons, with the exception of a few pentagons.
Errors for this case are computed considering only vertices or edges for which their latitude $\lambda$ satisfies $| \lambda | > 20^{\circ}$. This is because, as explained in \cite{turner2021mpas}, the MPAS-Seaice grid is rotated so that the poles of the eastward and northward directions are placed at the equator, to avoid a convergence of the northward components of the velocity at the geographic poles. Hence, with this rotation, the errors due to the metric terms will be prevalent at the equator where no sea ice is present, and therefore leaving these latitudes out of the calculation of the errors is justified.

We are assuming the following constitutive relation $\bm{\sigma}_{ij} = {\dot {\bm{ \epsilon}}}_{ij}$, hence the analytical divergence of the stress field that we use to compute errors is obtained using derivatives of the  velocity field $\bm{u} = (u,v)$ given by
\begin{equation}
    u(\lambda,\phi) = Y^3_5(\pi/2-\lambda,\phi), \quad  v(\lambda,\phi) = Y^2_4(\pi/2-\lambda,\phi),
\end{equation}
 where $(\lambda,\phi)$ are latitude and longitude, and $Y$ is a spherical harmonic function.
 Considering the metric terms \cite{hunke2002elastic} and geographical coordinates, the strain rate is given by \cite{turner2021mpas}
        \begin{equation}\label{eq:strainSphere}
     \dot \epsilon_{11} = \dfrac{1}{r\cos(\lambda)}\dfrac{\partial u}{\partial \phi} -\dfrac{v}{r}\tan(\lambda), \quad \dot \epsilon_{22} = \dfrac{1}{r}\dfrac{\partial v}{\partial \lambda}, \quad
     \dot \epsilon_{12} = \frac{1}{2r}\dfrac{\partial u}{\partial \lambda} + \frac{1}{2r\cos(\lambda)}\dfrac{\partial v}{\partial \phi} +\dfrac{u}{2r}\tan(\lambda).
      \end{equation}   
The stress divergence field $\bm{F}=(F_1,F_2)$ we use as an analytical solution is also expressed using geographical coordinates \cite{malvern1969introduction}, and is the same used in \cite{turner2021mpas} 
      \begin{equation}
      \begin{aligned}\label{eq:stressSphere}
       &F_1 = \frac{1}{r\cos(\lambda)}\dfrac{\partial \dot \epsilon_{11}}{\partial \phi} + \frac{1}{r}\dfrac{\partial \dot \epsilon_{12}}{\partial \lambda} - \frac{2}{r} \tan(\lambda)\dot \epsilon_{12},\\
       &F_2=\frac{1}{r\cos(\lambda)}\dfrac{\partial \dot \epsilon_{12}}{\partial \phi} + \frac{1}{r}\dfrac{\partial \dot \epsilon_{22}}{\partial \lambda} +\frac{1}{r}\tan(\lambda)(\dot \epsilon_{11}-\dot \epsilon_{22}).
       \end{aligned}
     \end{equation}
     We observe that, to have the B-grid and the CD-grid numerical solutions converge to the analytical expression above,  the values of the functions $C_i(r)$ with $i=1,2,3$ have to be different with the two methods. Namely, for the B-grid $C_1(r)=C_2(r)=1/r$ and $C_3(r)=0$,
     a choice that is consistent with the definition of the strain in Eq. \eqref{eq:strainSphere}.
     For the CD-grid, we have to set $C_1(r)=C_3(r)=1/r$ and $C_2(r)=2/r$, which means the strain rate is corrected once substituted into the discrete expression of the divergence of the stress, with extra terms that 
 match those multiplied by $\tan(\lambda)$ in Eq. \eqref{eq:stressSphere}. At the moment, we could not reach a definite conclusion on why this correction is necessary for the CD-grid.
 As shown in \cite{turner2021mpas}, the B-grid approach can converge on a spherical mesh with  $A_{P_e}$ being either the area of the dual triangle centered at a given vertex (standard approach), or the lumped mass matrix type of quantity defined in \eqref{eq:altDen} (consistent approach).
 Note that for the B-grid, the definition in  \eqref{eq:altDen} is different in that the integrals are over the three cells that own a given vertex.
 On the other hand, we found that the CD-grid approach on the sphere converges only if the choice in \eqref{eq:altDen} is considered for $A_{P_e}$ (consistent approach), i.e. the diamond-shape option does not provide convergence. 
 Therefore, for both the B-grid and the CD-grid we use 
 the choice of $A_{P_e}$ defined in \eqref{eq:altDen}, hence in the $L_2$ norm computation  in \eqref{eq:L2norm} the areas $A_i$ are equal to $A_{P_e}$ in \eqref{eq:altDen}.
 Another difference between the B-grid and the CD-grid approach lies in the way the matrices in Eq. \eqref{eq:matrices} and Eq. \eqref{eq:matrices2} are computed. Namely, for the B-grid, it is sufficient to project the vertex coordinates on a plane tangent to the sphere at the cell center.
 This means that, in general, for a given vertex, the contributions coming from the three cells that own it (see Figure \ref{fig:2} (left) ) would not be computed on the same plane.
 For the CD-grid, on the other hand, to ensure convergence it is necessary to project all the four shapes in Figure \ref{fig:2} (right) on the same tangent plane at an  edge location, so for a given edge the contributions coming from the four shapes are all computed on the same plane.
 
 Besides the $L_2$ norm defined in \eqref{eq:L2norm}, for the spherical tests we also consider the $L_{\infty}$ norm defined as
 \begin{equation}
     \max_i | F_i - \widetilde{F}_i |,
 \end{equation}
 where $F_i$ and $\widetilde{F}_i$ are as in \eqref{eq:L2norm} and the maximum is taken over all vertices or edges that take part in the computation, depending on whether a B-grid or a CD-grid approach is used.
 
\begin{figure}[!t]
   \centering
      \includegraphics[scale=0.55]{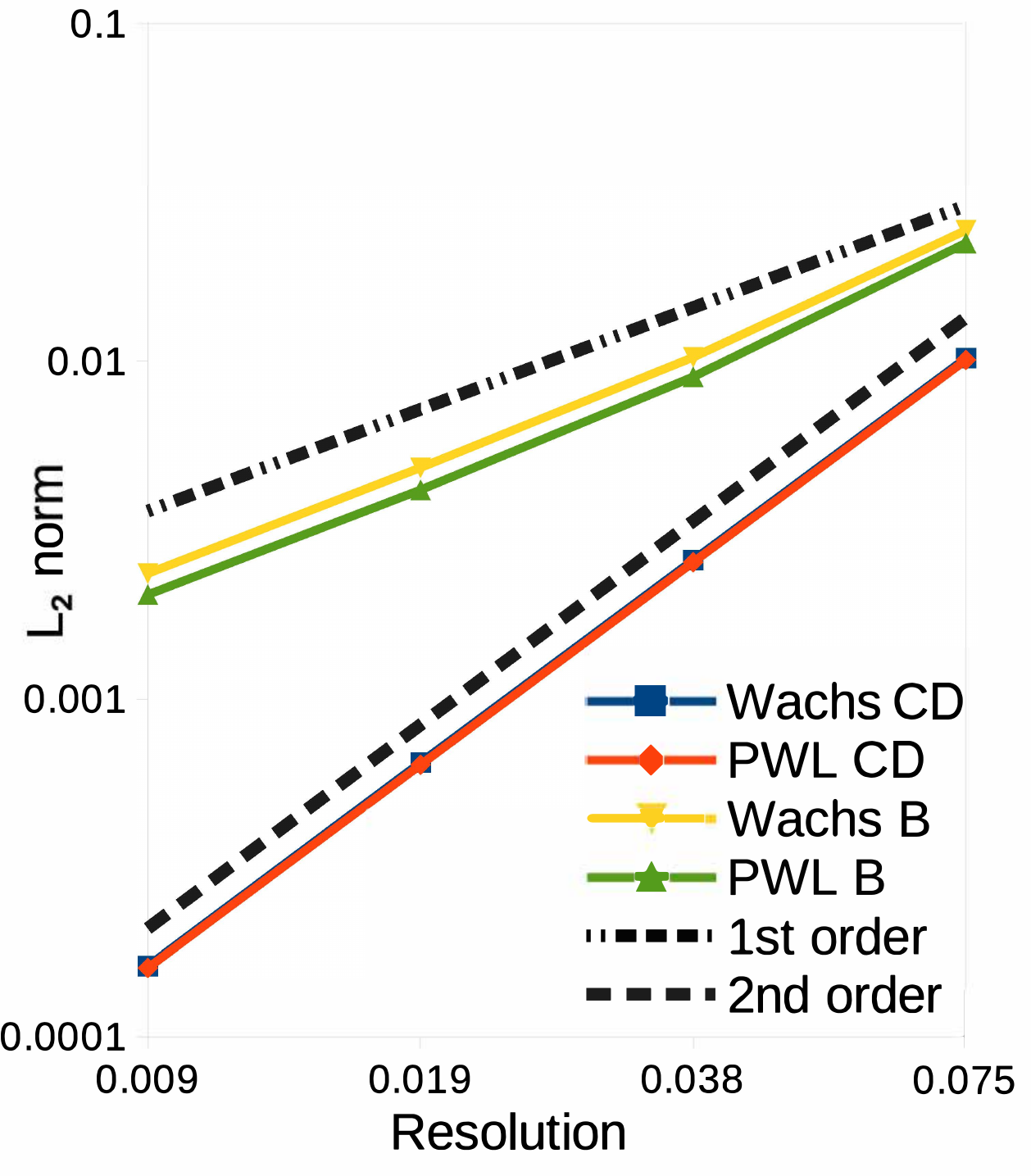}
   \includegraphics[scale=0.55]{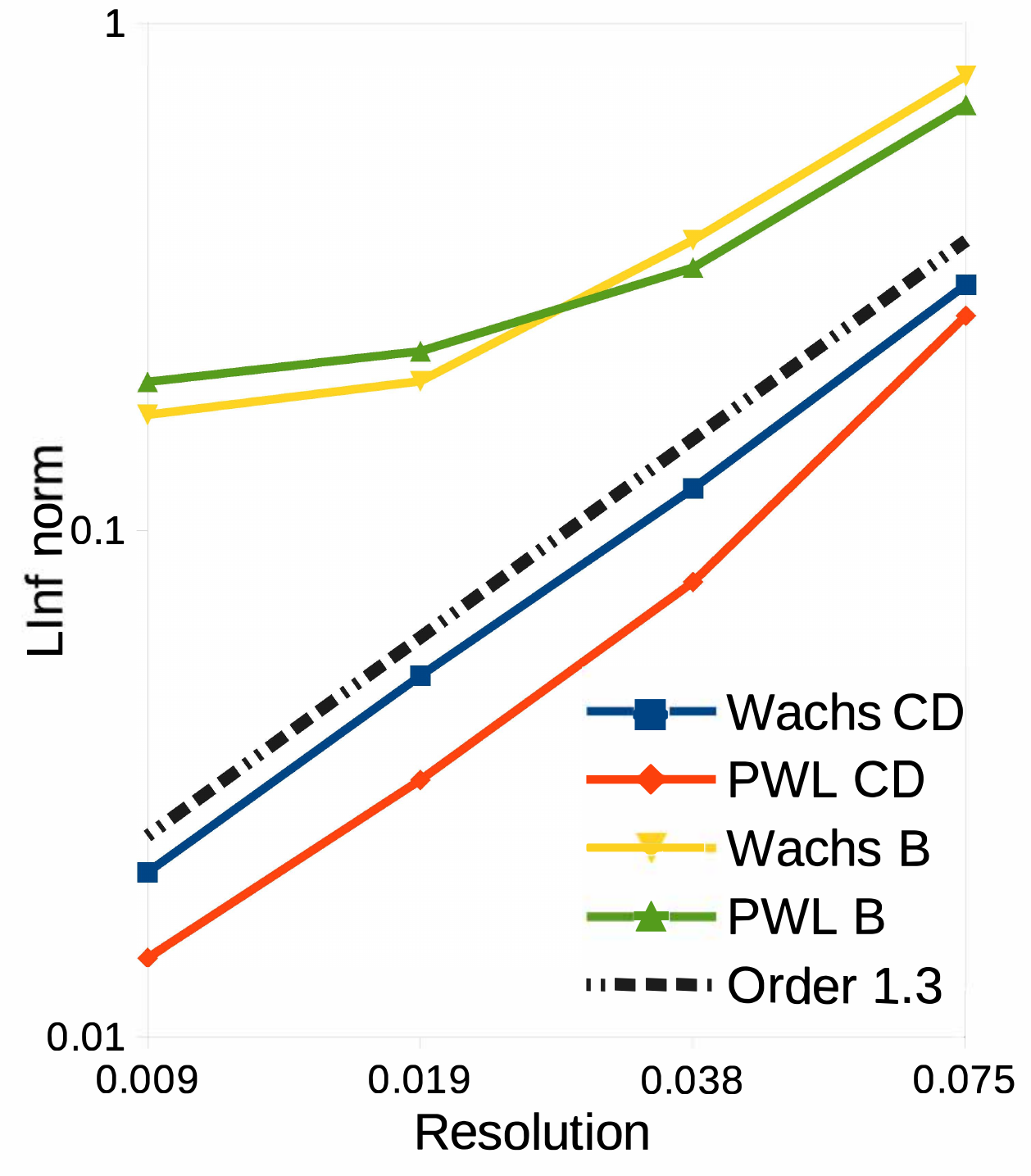}
\caption{Convergence on the unit sphere for the eastward component of the divergence of the stress.
Left: $L_2$ norm. Right: $L_{\infty}$ norm. Wachs: Wachspress basis. PWL: piecewise linear basis.}
   \label{fig:sphereConv}
\end{figure}

\begin{figure}[!t]
   \centering
      \includegraphics[scale=0.55]{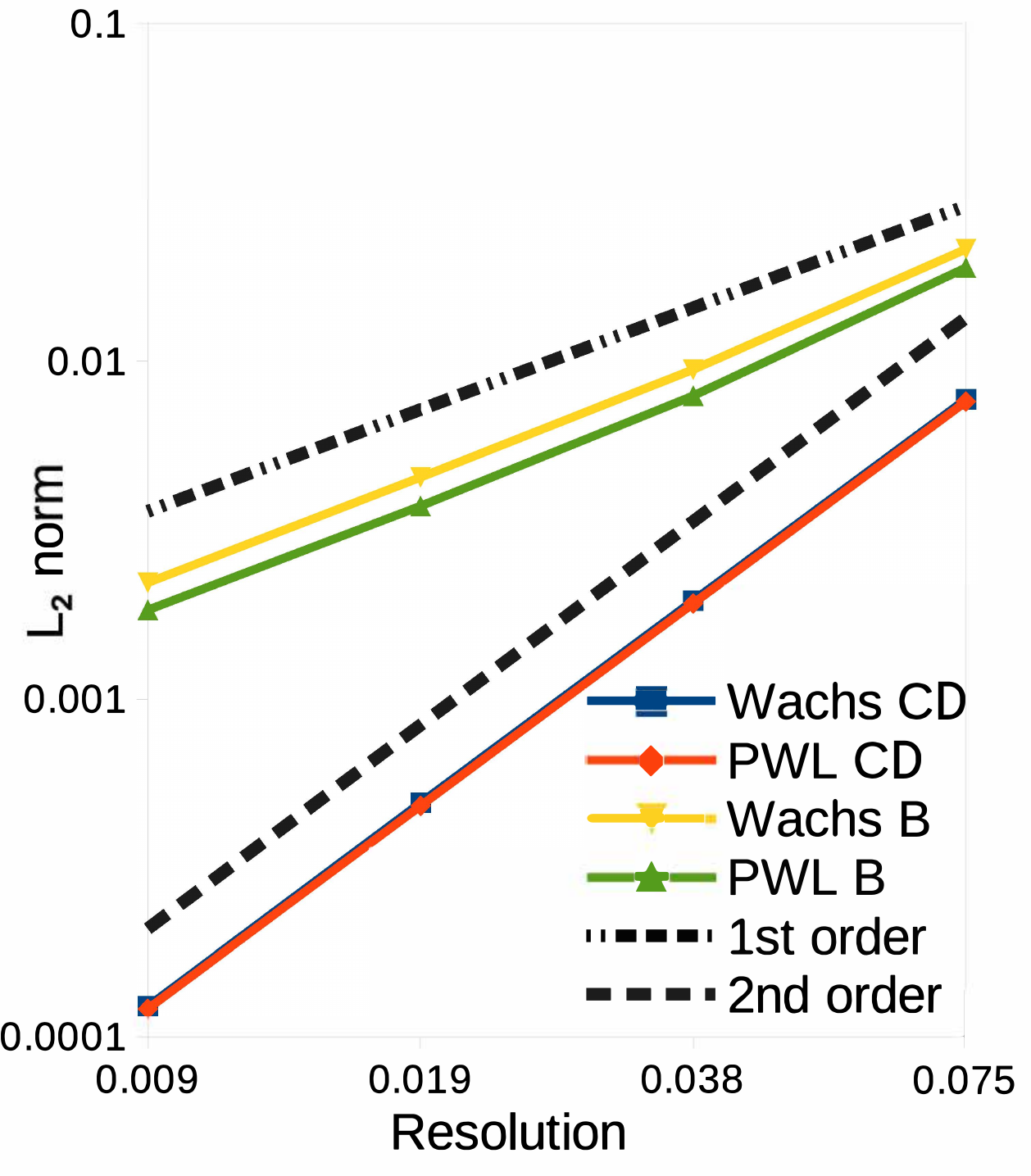}
   \includegraphics[scale=0.55]{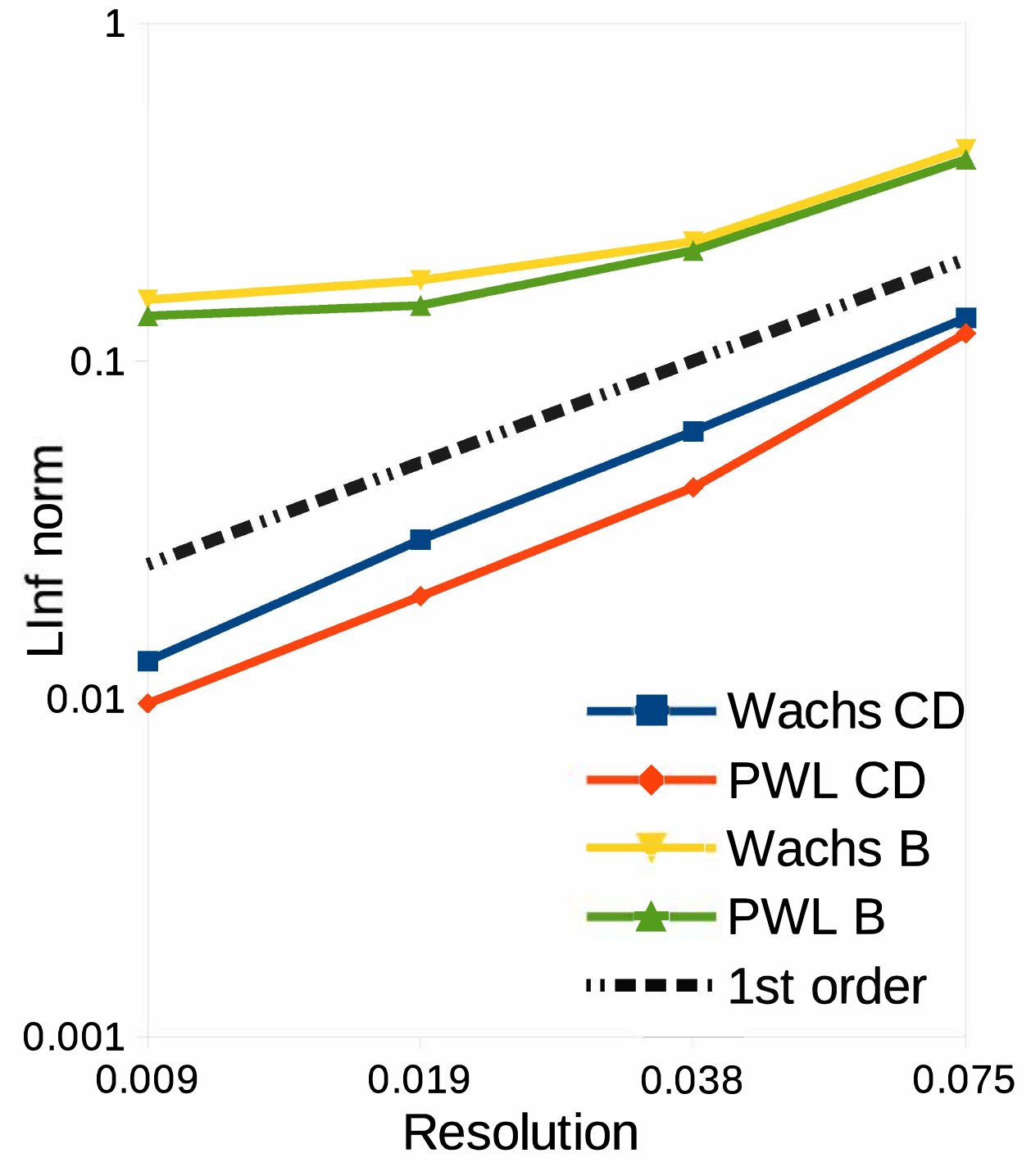}
\caption{Convergence on the unit sphere for the northward component of the divergence of the stress.
Left: $L_2$ norm. Right: $L_{\infty}$ norm. Wachs: Wachspress basis. PWL: piecewise linear basis.}
   \label{fig:sphereConv2}
\end{figure}

\begin{figure}[!t]
   \centering
      \includegraphics[scale=0.295]{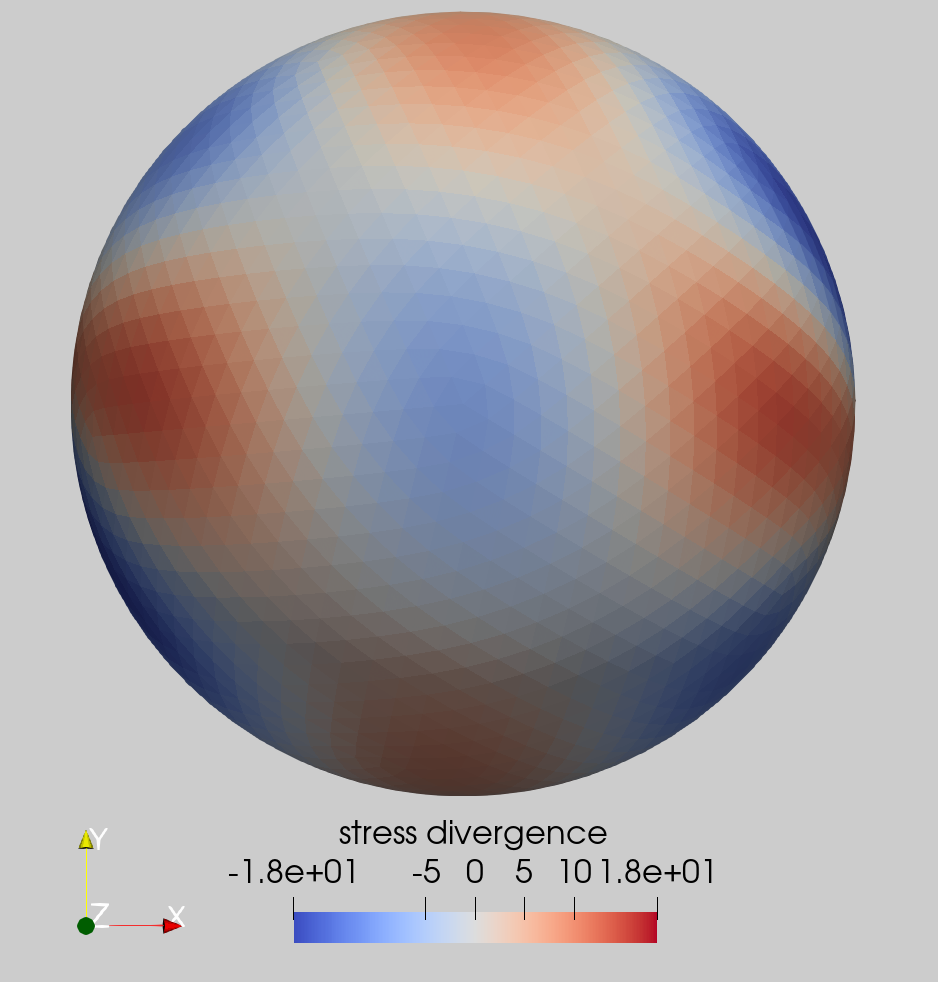}
  \includegraphics[scale=0.3]{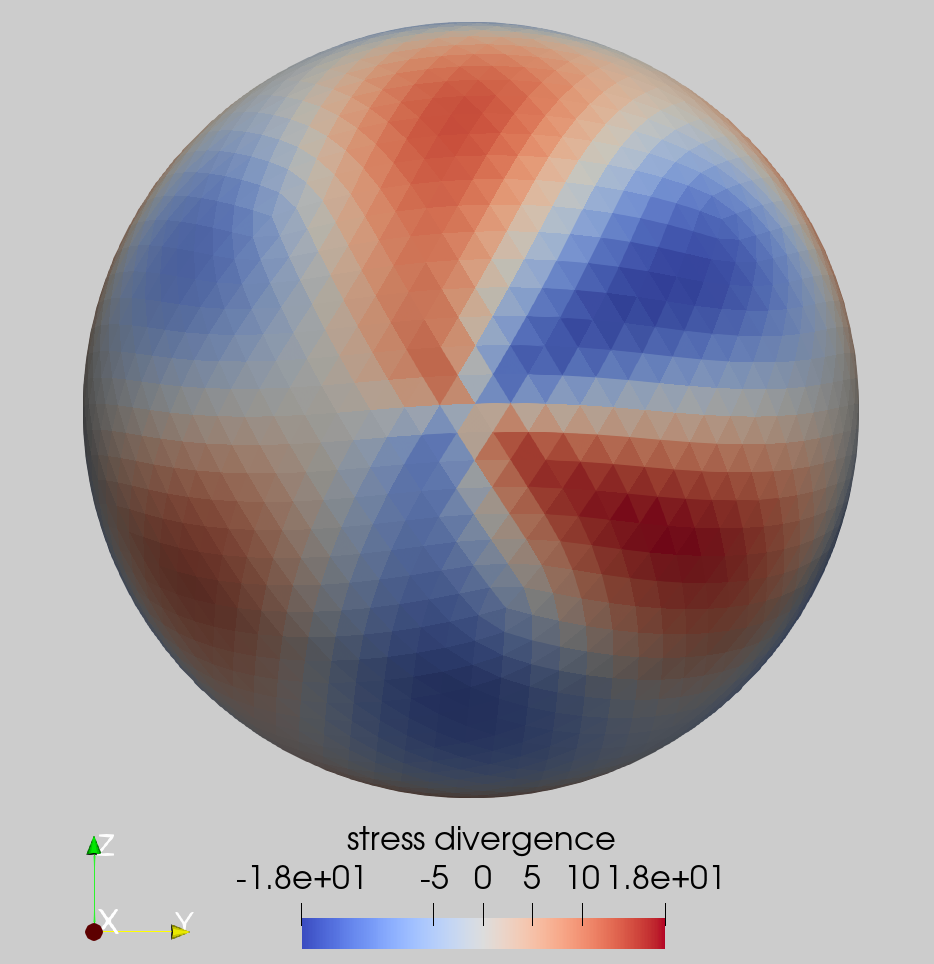}
    \includegraphics[scale=0.298]{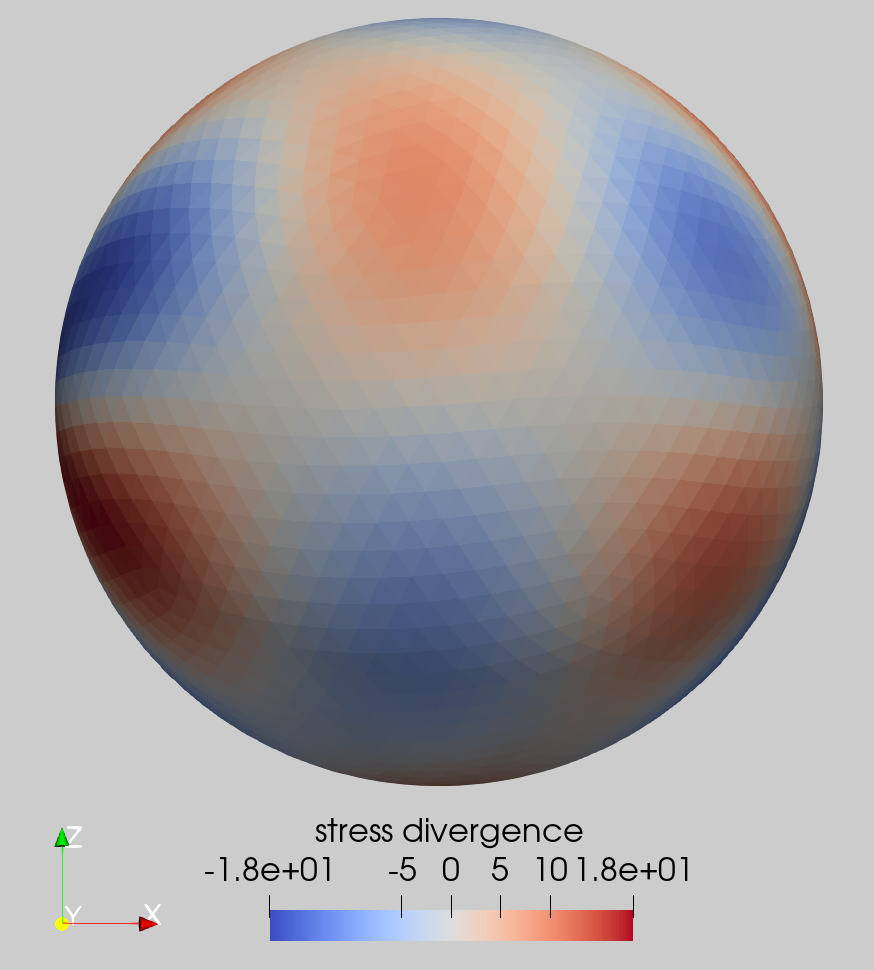}\\
    \vspace{0.5cm}
         \,\includegraphics[scale=0.3]{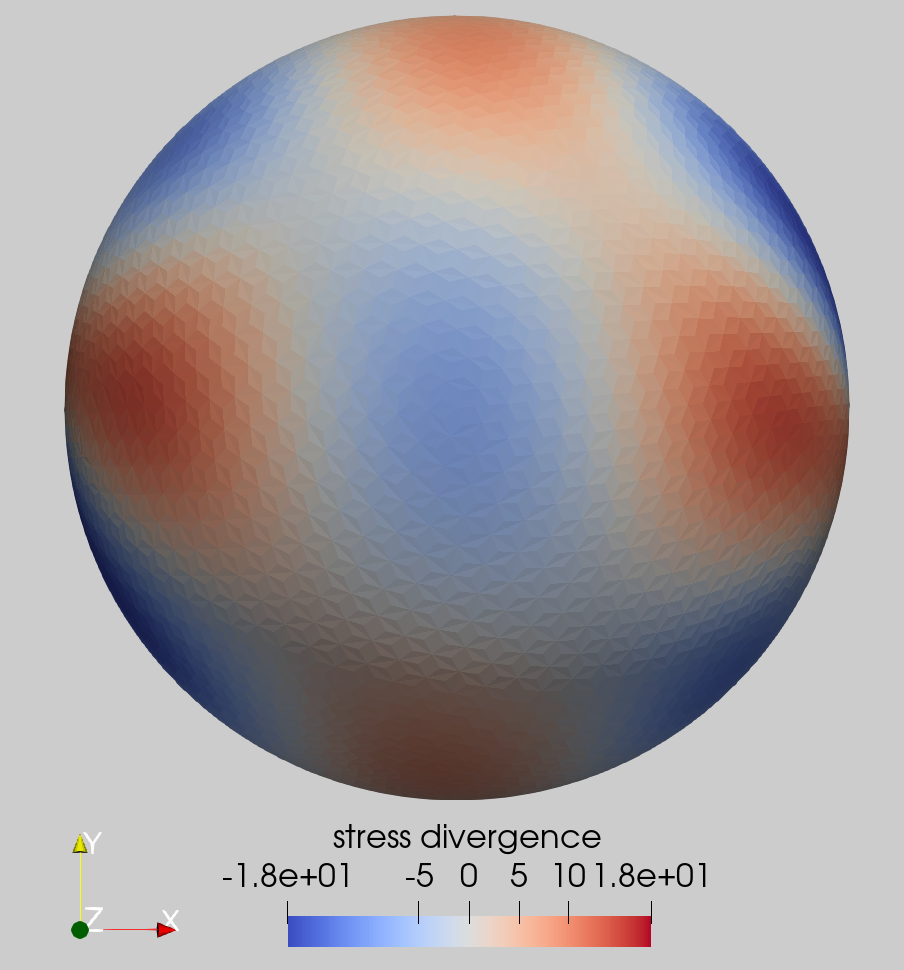}
  \includegraphics[scale=0.302]{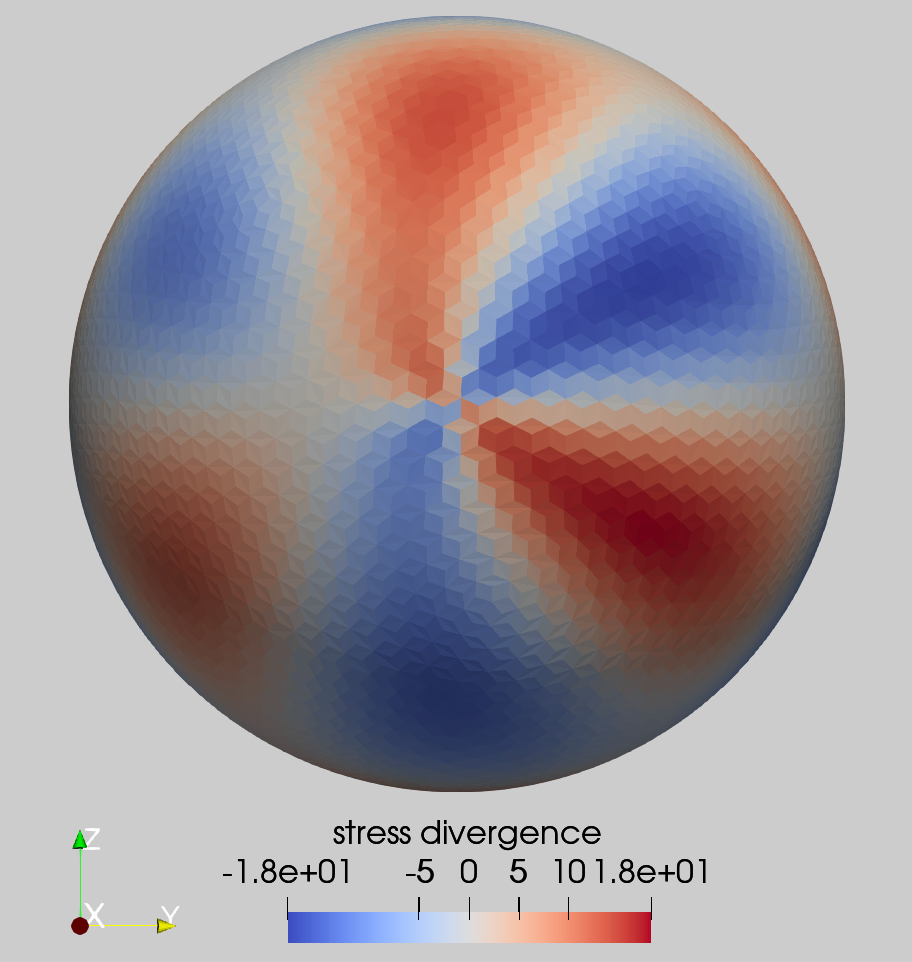}\,
    \includegraphics[scale=0.292]{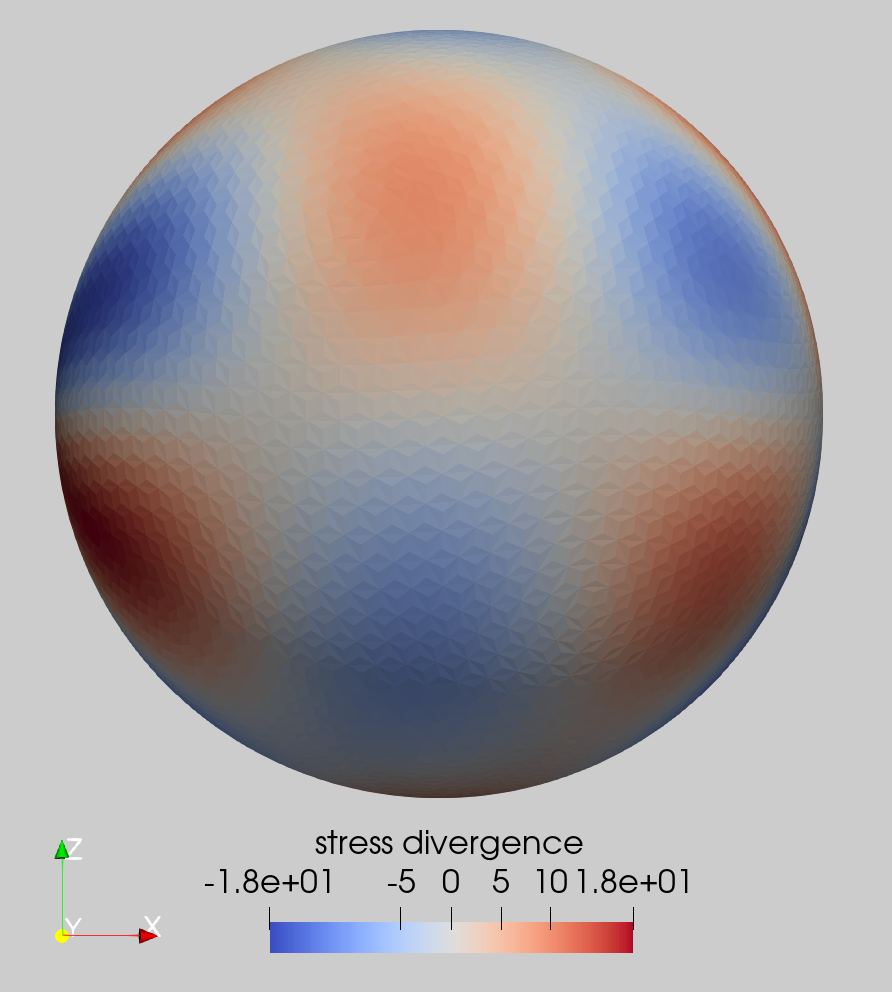}
\caption{Numerical solution for the eastward component of the divergence of the stress. Top: B-grid approach. Bottom: CD-grid approach. All pictures consider PWL basis functions and lowest resolution.}
   \label{fig:paraview}
\end{figure}

\begin{figure}[!t]
   \centering
      \includegraphics[scale=0.405]{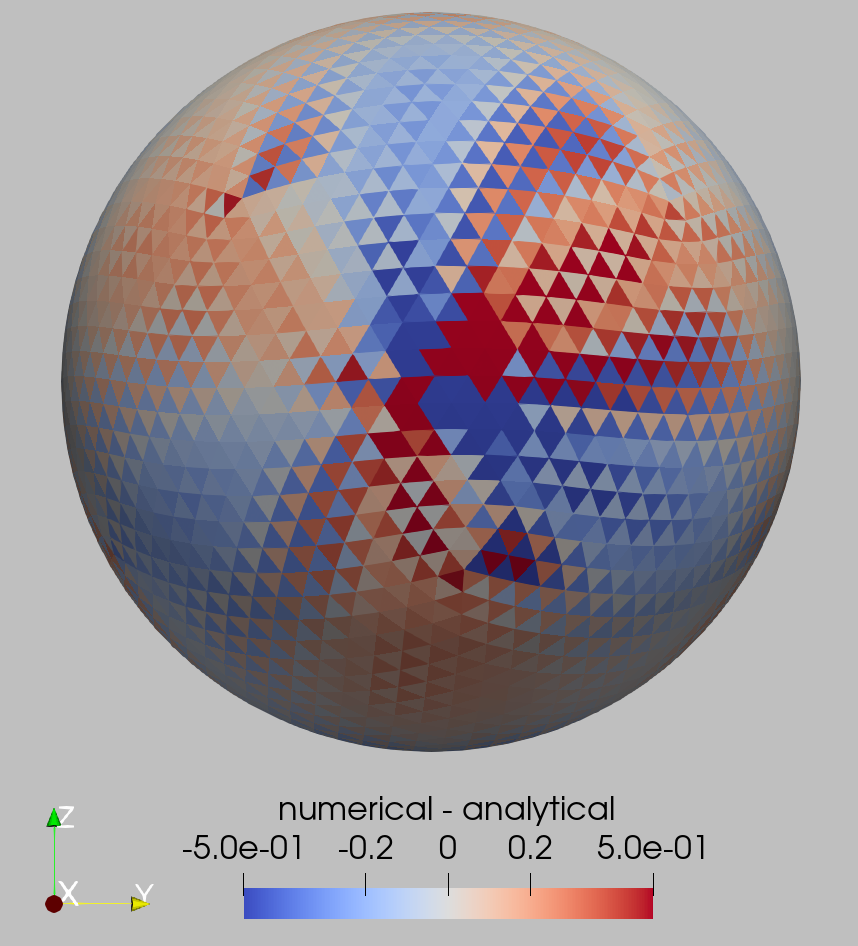}
  \includegraphics[scale=0.401]{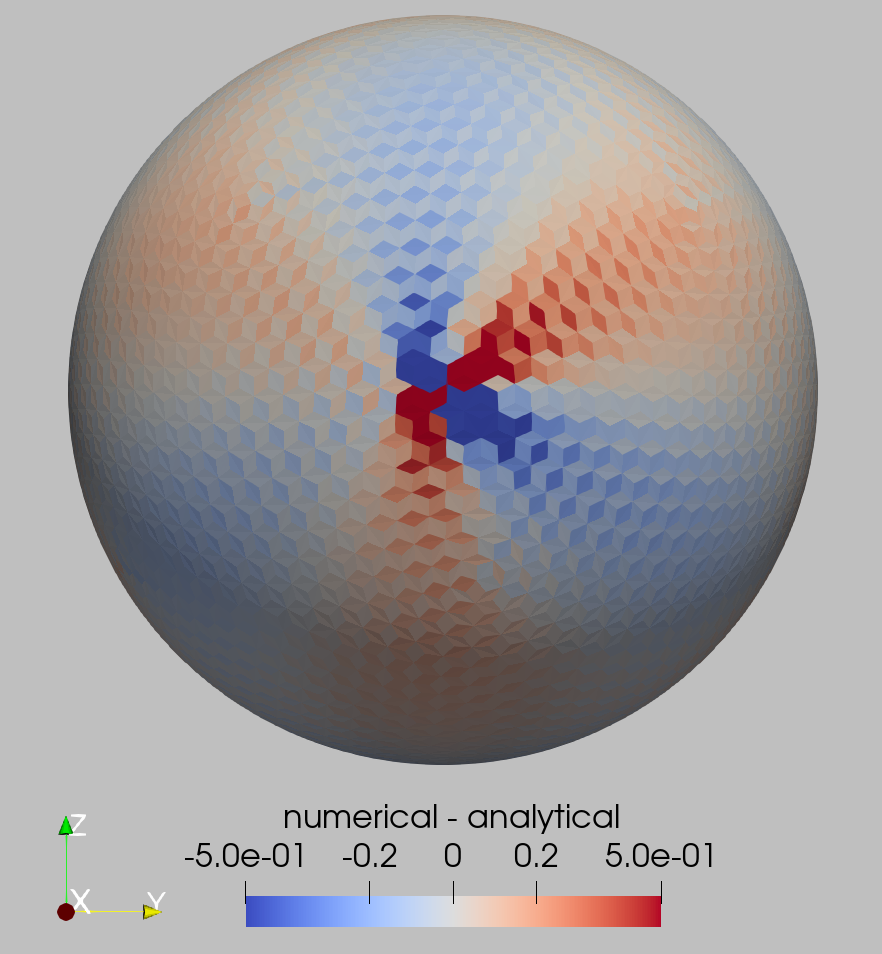}
\caption{Left: error for the B-grid approach. Right: error for the CD-grid approach. Both pictures consider the eastward component of the divergence of the stress, PWL basis functions and lowest resolution of the grid. }
   \label{fig:paraErrors}
\end{figure}

\begin{figure}[!t]
   \centering
      \includegraphics[scale=0.41]{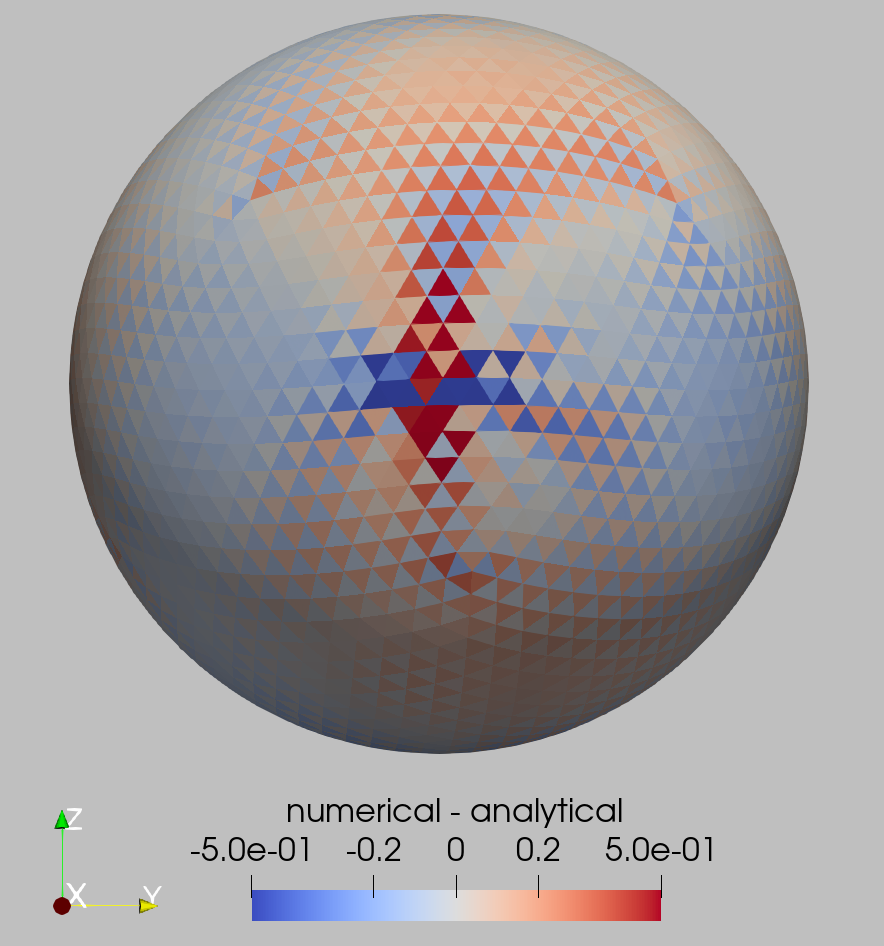}
  \includegraphics[scale=0.4075]{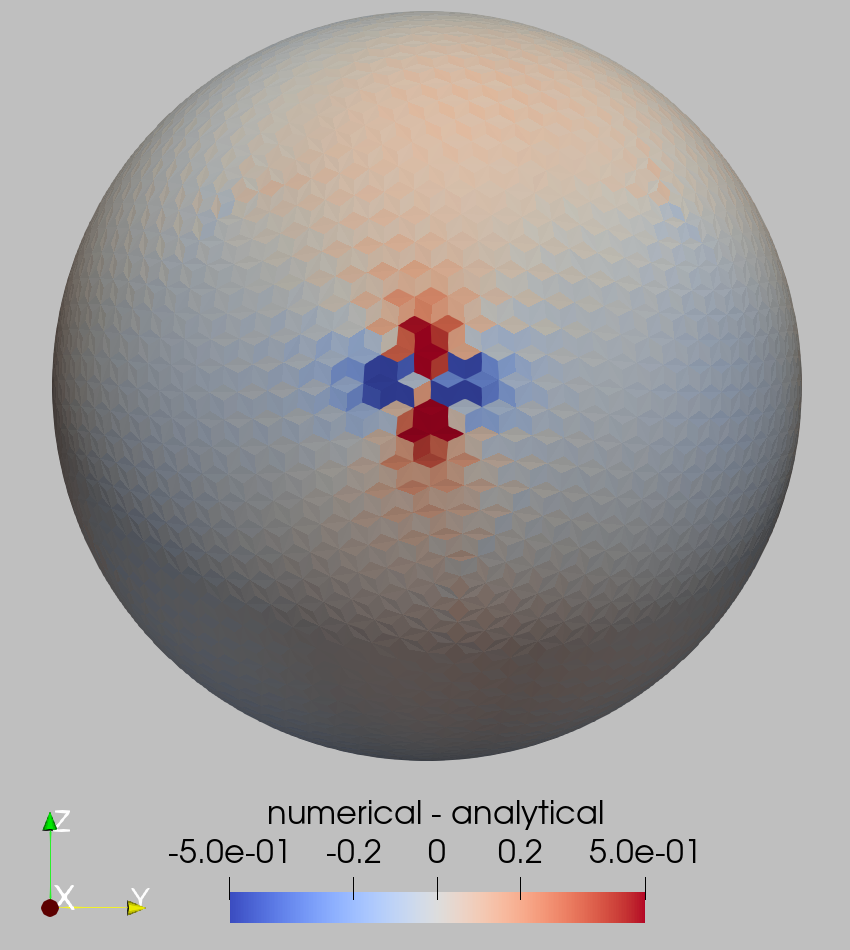}
\caption{Left: error for the B-grid approach. Right: error for the CD-grid approach. Both pictures consider the northward component of the divergence of the stress, PWL basis functions and lowest resolution of the grid. }
   \label{fig:paraErrors2}
\end{figure}

 Results are shown in Figure \ref{fig:sphereConv} for the case of the eastward component of the divergence of the stress and in Figure \ref{fig:sphereConv2} for the northward.  
 %Let us split $\dfrac{\partial D_2}{\partial u_{u e}}$ in Eq. \eqref{eq:D2_superFinal} as the sum of two contributions
 %\begin{equation}\label{eq:D2decomp}
%     \dfrac{\partial D_2}{\partial u_{u e}} = \Big(\dfrac{\partial D_2}{\partial u_{u e}}\Big)_{NM} + C_2(r)\Big(\dfrac{\partial D_2}{\partial u_{u e}}\Big)_{M},
 %\end{equation}
 %where the first term on the right hand side takes into account the non-metric terms and the second represents the metric terms.
 %Then, in Figure \ref{fig:parComparison}, we compare  $\Big(\dfrac{\partial D_2}{\partial u_{u e}}\Big)_{M}$ for the B-grid (right) and C-grid (left), and notice that the term is qualitatively the same for both methods. This suggests than the need to have $C_2(r)=2/r$ for the C-grid does not come from an underestimation of the metric terms $\Big(\dfrac{\partial D_2}{\partial u_{u e}}\Big)_{M}$ but rather from the need to correct more substantially the non metric terms in $\Big(\dfrac{\partial D_2}{\partial u_{u e}}\Big)_{NM}$.
  %%%%%%%%%%%%
%\begin{figure}[!t]
%   \centering
%      \includegraphics[scale=0.396]{figures/metricC.png}
%   \includegraphics[scale=0.40]{figures/metricB.png}
%\caption{Metric terms in Eq. \eqref{eq:D2decomp}. Left: C-grid. Right: B-grid. Plots refer to the PWL choice of basis functions and lowest grid resolution.}
%   \label{fig:parComparison}
%\end{figure}
%%%%%%%%%%%%
  %%%%%%%%%%%%
%%%%%%%%%%%%
   %%%%%%%%%%%%
%%%%%%%%%%%%
We observe from Figure \ref{fig:sphereConv} that while the B-grid approach shows a first-order convergence rate for the $L_2$ norm, the CD-grid  remains second-order as it was on a planar mesh. Moreover, the B-grid does not convergence in the $L_{\infty}$ norm for either choice of basis function whereas the CD-grid approach shows a convergence rate that is slightly better than linear, with the PWL choice showing lower errors than the Wachspress.
For the northward component, the results in Figure  \ref{fig:sphereConv2} show a similar behavior.
In Figure \ref{fig:paraview}, we are displaying the eastward component of the divergence of the stress for the B-grid and the CD-grid considering different views. The images refer to the case of PWL basis functions and the lowest resolution considered.
  %%%%%%%%%%%%
%%%%%%%%%%%%
 In Figure \ref{fig:paraErrors} we are showing the errors of the numerical solutions compared to the analytical for the eastward component of the divergence of the stress. Notice how, as expected, the higher values of the error are in proximity of latitudes that are excluded from the computation of the errors. A similar comparison for the northward component is in Figure  \ref{fig:paraErrors2}.
   %%%%%%%%%%%%
%%%%%%%%%%%%
   %%%%%%%%%%%%
%%%%%%%%%%%%

%%%%%%%%%%%%

\subsection{Velocity solver in a square domain}
 
\begin{figure}[!t]
   \centering
      \includegraphics[scale=0.32]{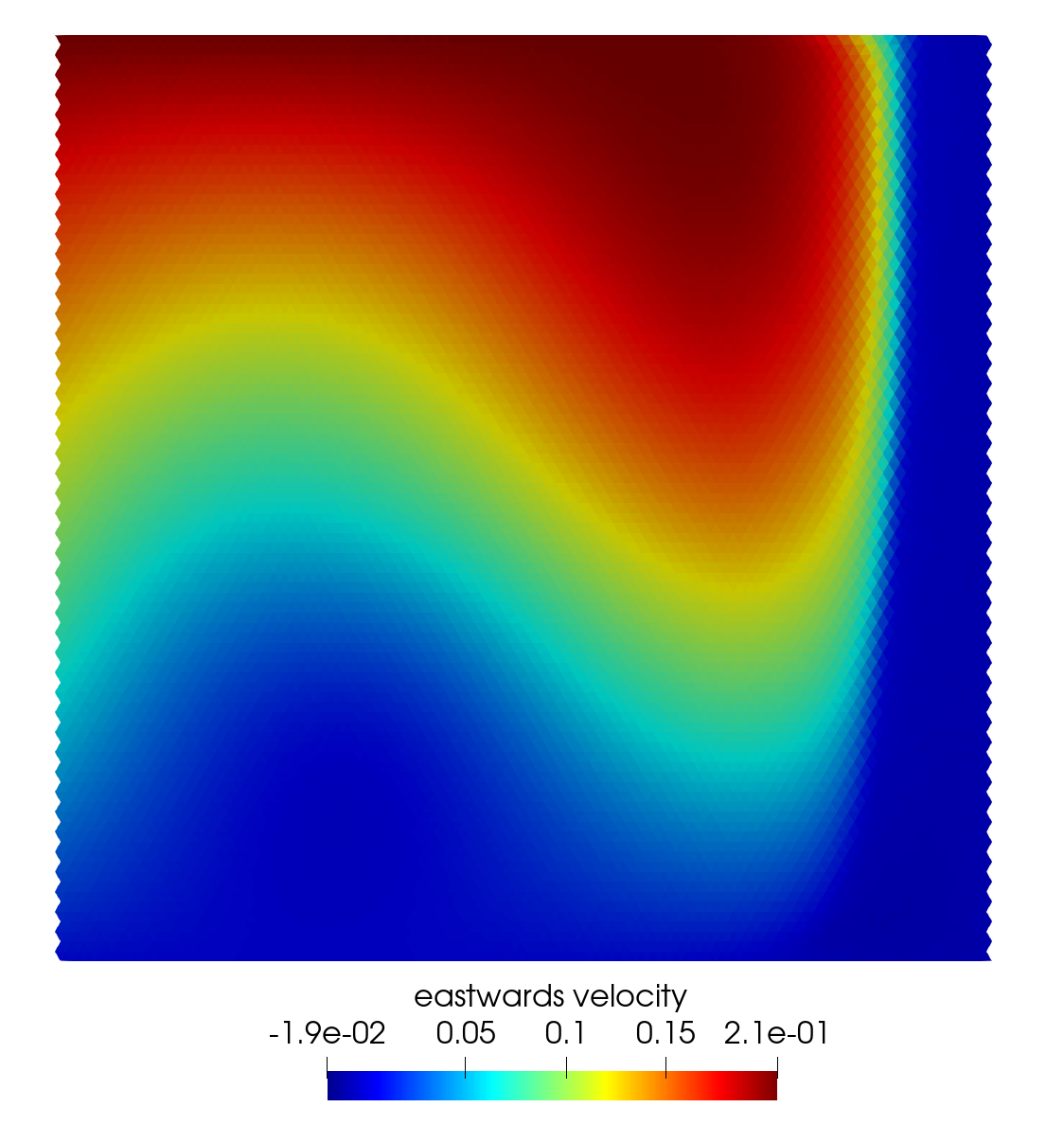}
  \includegraphics[scale=0.3145]{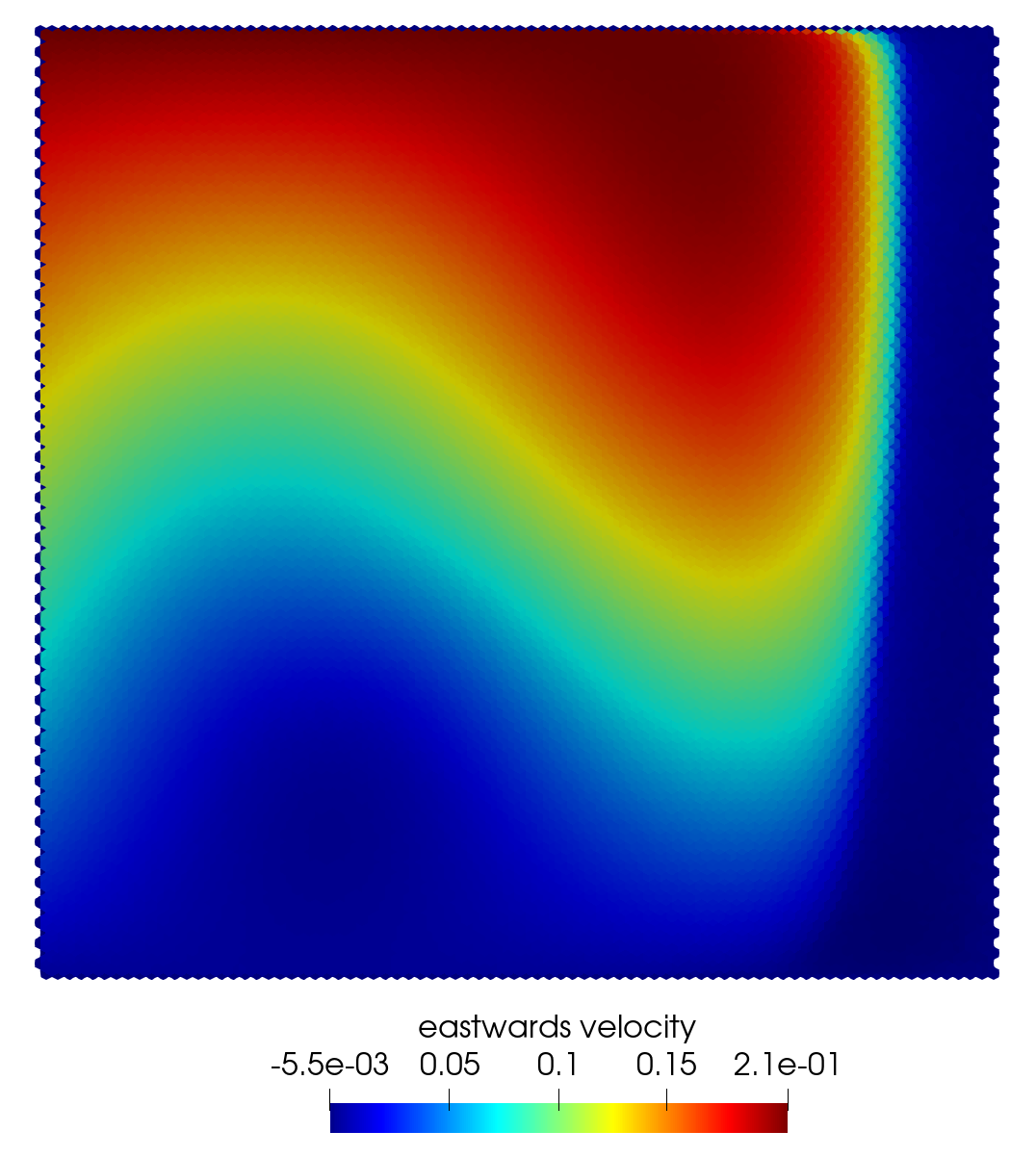}
\caption{Left: eastward velocity for the B-grid. Right: eastward velocity for the CD-grid. Both pictures refer to Wachspress basis functions and the final time of the simulation.}
   \label{fig:parVels1}
\end{figure}

\begin{figure}[!t]
   \centering
      \includegraphics[scale=0.32]{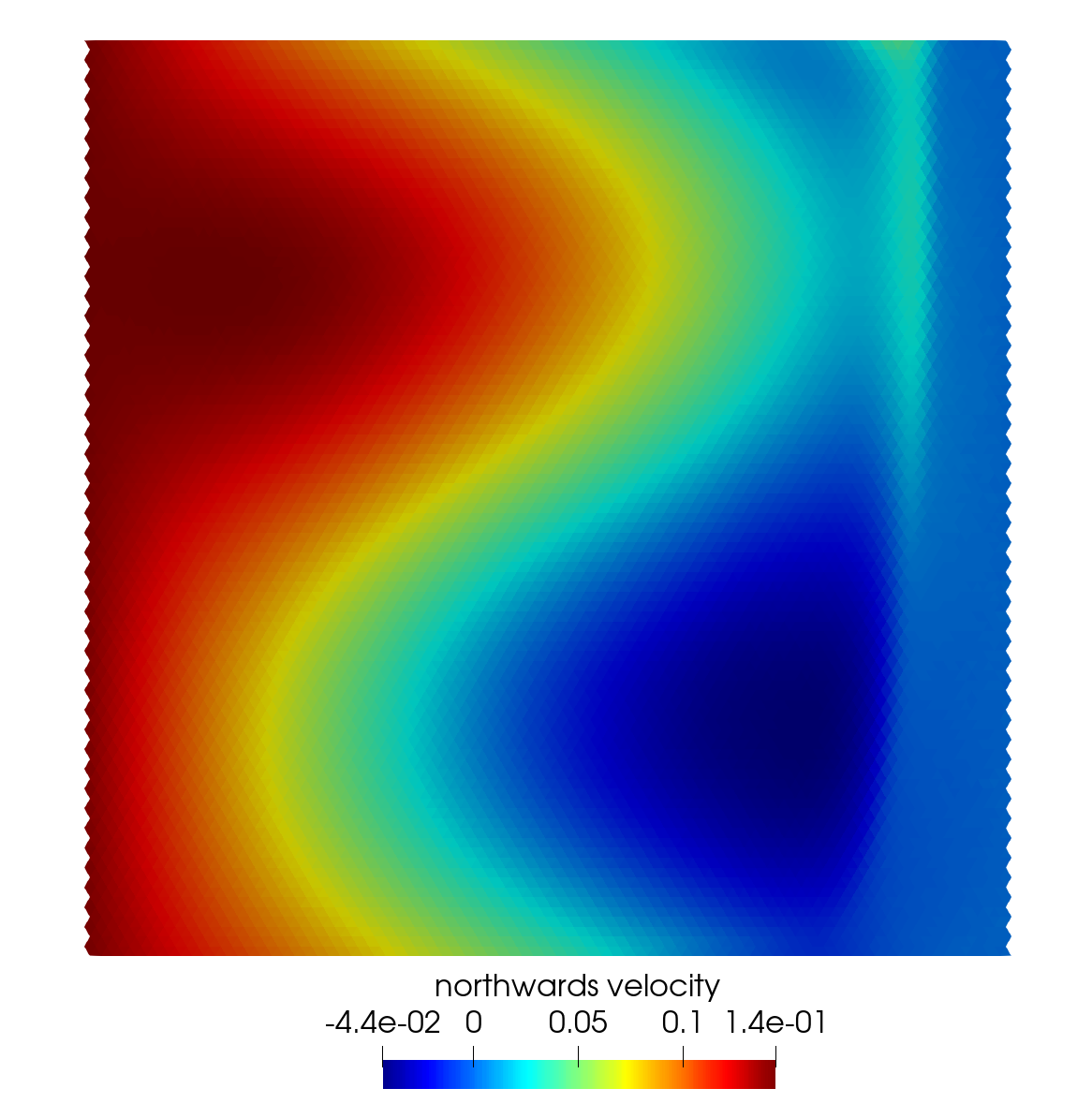}
  \includegraphics[scale=0.32]{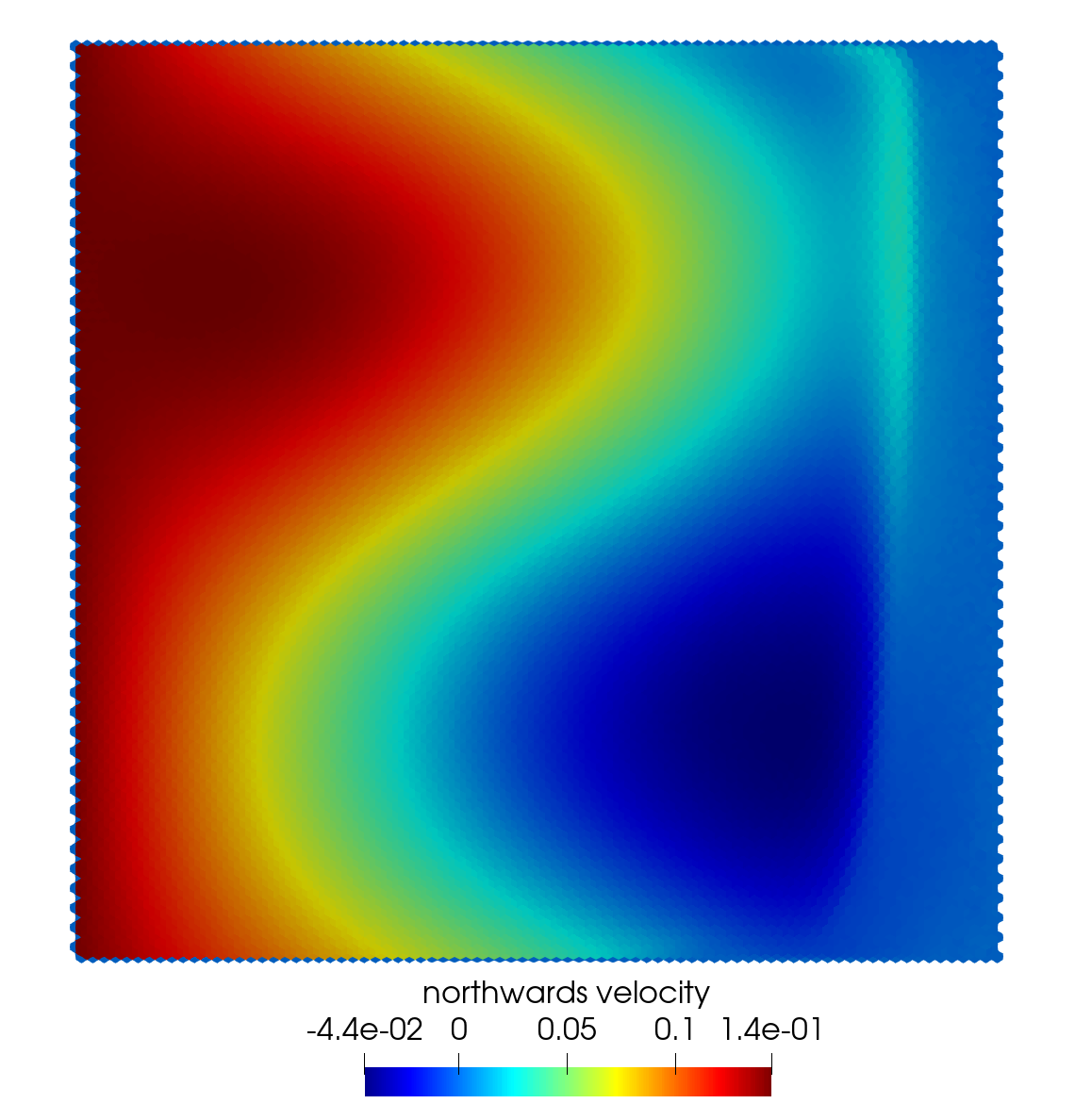}
\caption{Left: northward velocity for the B-grid. Right: northward velocity for the CD-grid. Both pictures refer to Wachspress basis functions and the final time of the simulation.}
   \label{fig:parVels2}
\end{figure}

\begin{figure}[!t]
   \centering
      \includegraphics[scale=0.33]{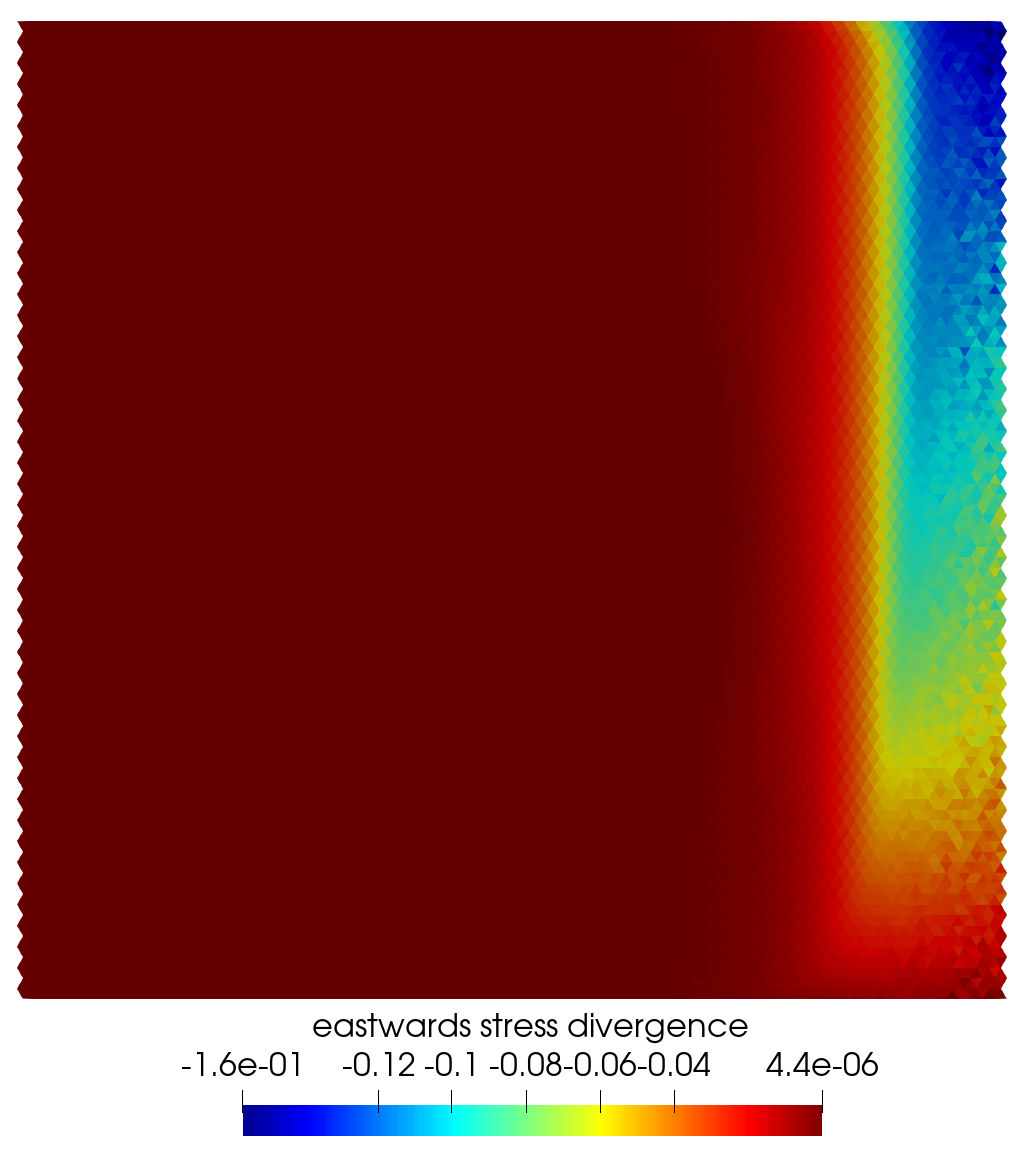}
  \includegraphics[scale=0.3245]{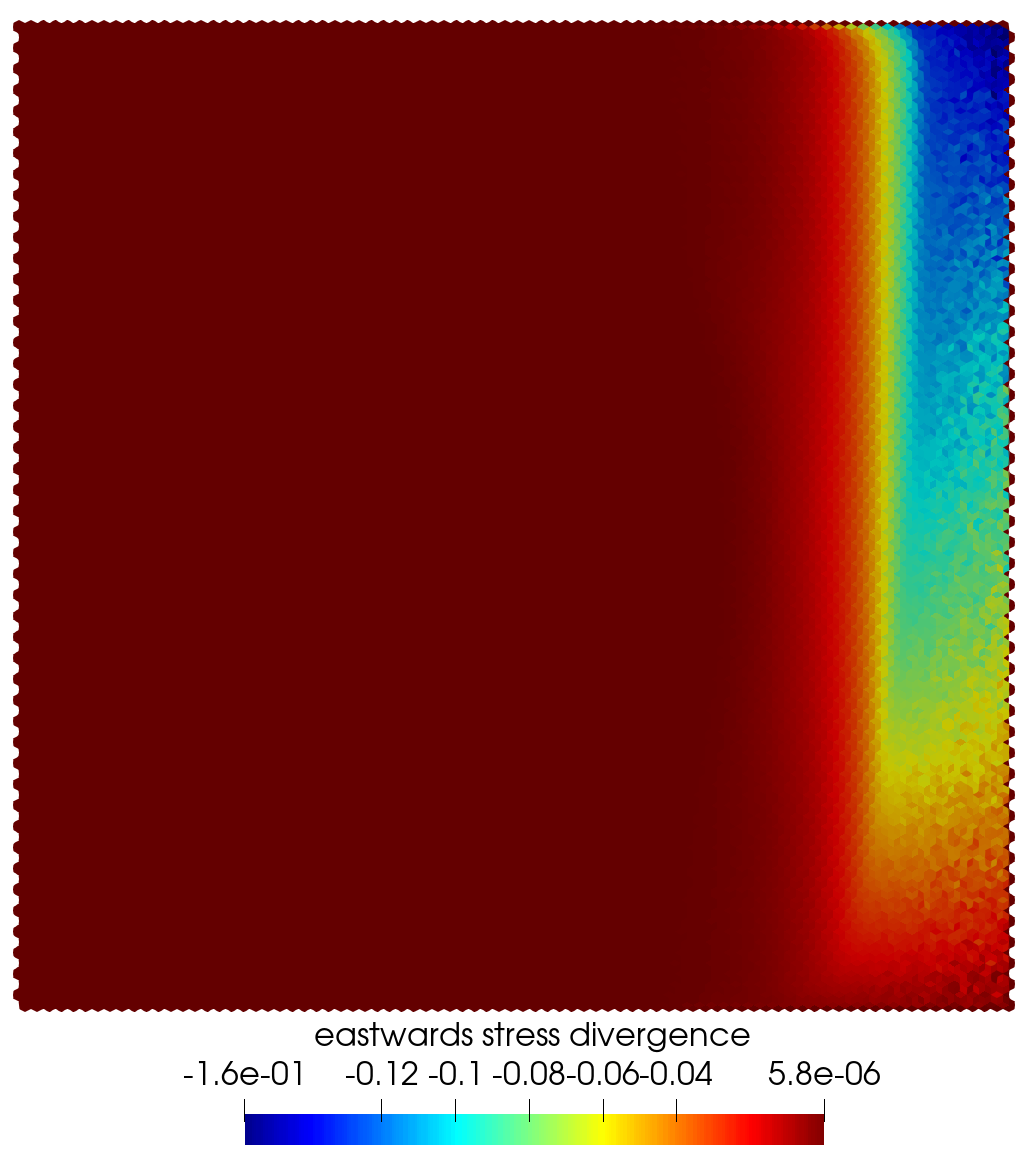}
\caption{Left: eastward divergence of the stress for the B-grid. Right: eastward divergence of the stress for the CD-grid. Both pictures refer to Wachspress basis functions and the final time of the simulation.}
   \label{fig:parStress1}
\end{figure}

\begin{figure}[!t]
   \centering
      \includegraphics[scale=0.325]{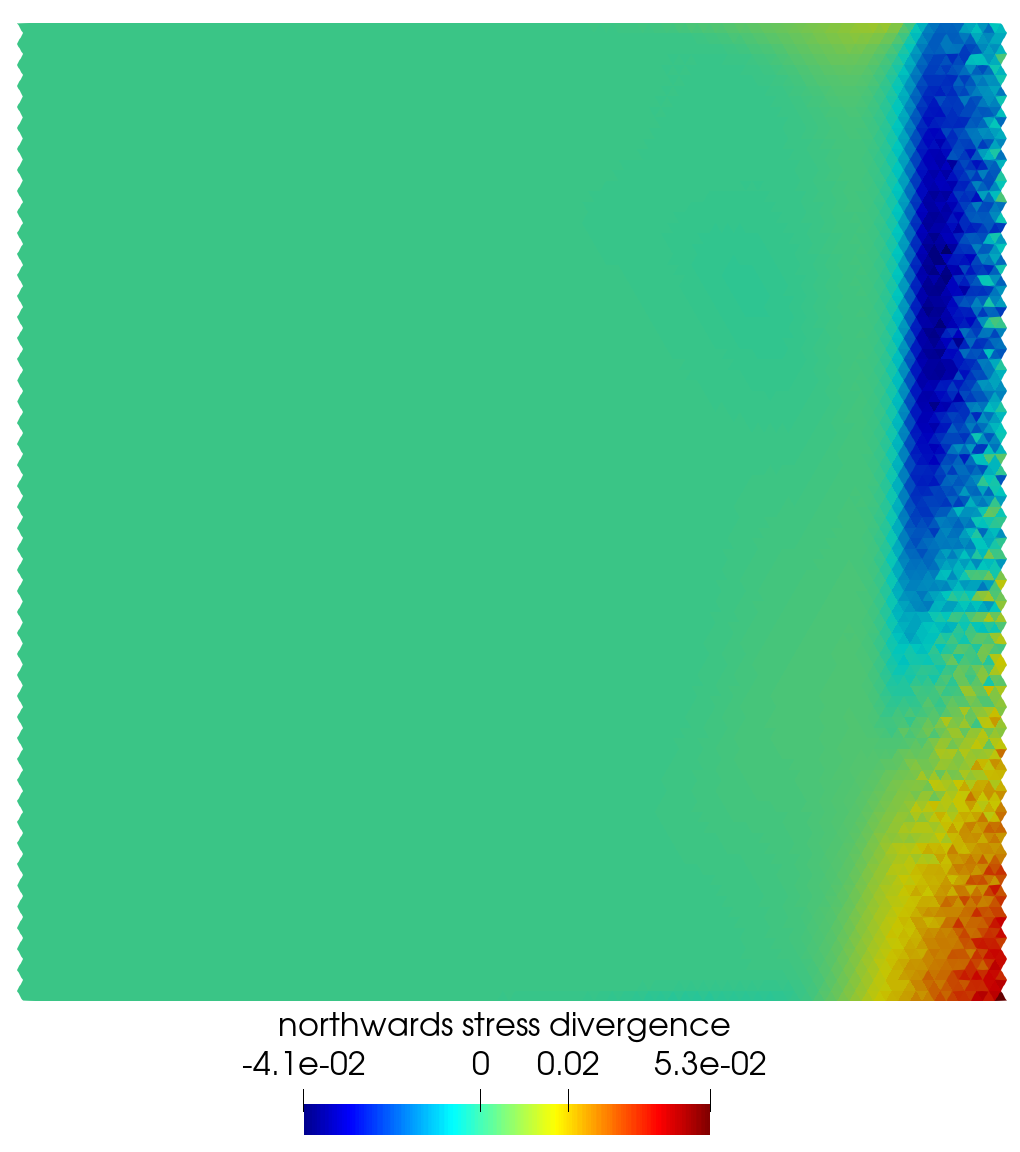}
  \includegraphics[scale=0.325]{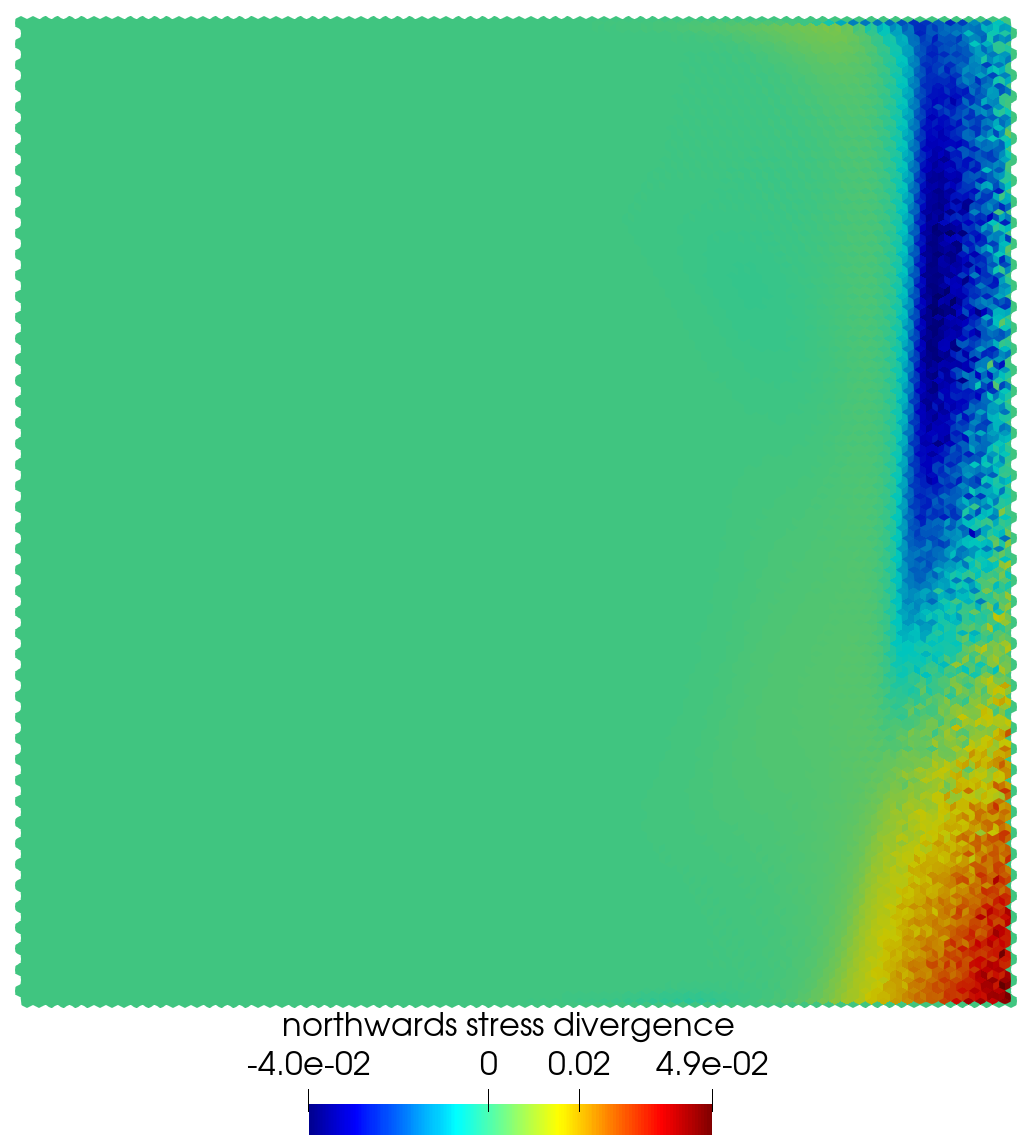}
\caption{Left: northward divergence of the stress for the B-grid. Right: northward divergence of the stress for the CD-grid. Both pictures refer to Wachspress basis functions and the final time of the simulation.}
   \label{fig:parStress2}
\end{figure}

 We conclude with some qualitative results obtained on a square domain of size 80 km as in \cite{turner2021mpas}, considering only the velocity solver and turning off advection and column physics.
 The setup is based on a similar test from \cite{hunke2001viscous}. The constitutive relation for this case is the EVP.
 No snow is present and ice thickness is fixed at 2 m. Ice concentration increases linearly in the eastward direction from zero at the western boundary to one at the eastern
boundary.  For the Coriolis term, we consider a constant $f$-plane with $f=$ 
1.46e-4 s$^{-1}$.
 Forcing terms originate from atmospheric winds $\bm{u}^a = (u^a,v^a)$ and ocean currents $\bm{u}^o= (u^o,v^o)$ of the form
 \begin{equation}
 \begin{aligned}
     &u^a = 5 - 3\sin(2\pi x /L_x)\sin(\pi y /L_y), \quad v^a = 5 - 3\sin(2\pi y /L_y)\sin(\pi x /L_x),\\
    & \qquad \qquad u^o =  0.1 ((2y-L_y)/L_y), \qquad v^o =  -0.1 ((2x-L_x)/L_x),
     \end{aligned}
 \end{equation}
 with $L_x$ and $L_y$ being the domain size in the eastward and northward directions respectively.
 The velocity solver is advanced with four time steps, and $\Delta t = 60$ min. The aim of this test is to show that the results obtained with the CD-grid are qualitatively similar to those obtained with the B-grid.
 For this purpose, we display in Figure \ref{fig:parVels1} and Figure \ref{fig:parVels2} the two velocity components for the two approaches, considering Wachspress basis functions and the final time of the simulation.
    %%%%%%%%%%%%
%%%%%%%%%%%%
We observe that both velocity profiles are very similar qualitatively, with only minimal differences at the top right corner of the domain.
For completeness, we also show the two components of the divergence of the stress in Figure \ref{fig:parStress1} and Figure \ref{fig:parStress2}.
 %%%%%%%%%%%%
 %%%%%%%%%%%%
 
 \section{Conclusions}\label{sec:concl}
We presented a promising new unstructured variational formulation on a CD-grid for the sea ice dynamics, focusing our analysis on the accuracy of the proposed method in approximating the divergence of the internal stress, which is arguably the most challenging term to discretize for the sea ice dynamics.
 Studying the convergence rate, we have shown that the proposed method is second-order accurate on a planar domain as well as in a spherical one and that is capable of reproducing similar results as the current B-grid formulation in MPAS-Seaice when used within a full velocity solver.
 More investigation on the method is needed to fully understand its inner workings, especially concerning the need for additional metric terms on the spherical domain, that seem to be necessary to achieve convergence to an analytical solution. Despite the necessity of further work, the method showed appealing features and proved to be more accurate than the current discretization in MPAS-Seaice, making it a viable alternative to be explored in the future.
 
 \section*{Acknowledgments} 
 The authors would like to thank Sara Calandrini, Darren Engwirda and Elizabeth Hunke for helpful discussions. GC was sponsored by the Center For Nonlinear Studies (CNLS) at Los Alamos Laboratory. MRP, AKT, and AFR were supported as part of the Energy Exascale Earth System Model (E3SM) project and the Integrated Coastal Modeling (ICoM) project, funded by the U.S. Department of Energy, Office of Science, Office of Biological and Environmental Research. This research used resources provided by the Los Alamos National Laboratory Institutional Computing Program, which is supported by the U.S. Department of Energy National Nuclear Security Administration under Contract No. 89233218CNA000001.

 \bibliography{manuscript.bib}
 \bibliographystyle{plain}
 
 \section*{Appendix A}\label{appA}
  \Giacomo{The integral in \eqref{eq:D2} can be expressed as
\begin{equation}\label{eq:D2bis}
 D_{u,e}^2  = -\sum\limits_{i=1}^{2}\int_{T_e^i} \dfrac{\partial}{\partial u}\Big(\sigma_{12} \Big[\dfrac{\partial u} {\partial y} + \dfrac{\partial v}{\partial x} + u \, C_2(r) \tan(\lambda)\Big] \Big)dA   -\sum\limits_{i=1}^{2}\int_{\widehat{V}_e^i}\dfrac{\partial}{\partial u}\Big( \sigma_{12} \Big[\dfrac{\partial u} {\partial y} + \dfrac{\partial v}{\partial x} + u \, C_2(r)\tan(\lambda)\Big]\Big) dA.
 \end{equation}
 Substituting the basis expansions, the above equation becomes
 \begin{equation}
 \begin{aligned}\label{eq:D2_der_ref}
      D_{u,e}^2  = &-\sum\limits_{i=1}^{2}\int_{T_e^i}  \dfrac{\partial}{\partial u_e}\Big(    \sum\limits_{j=1}^{n_t} {\sigma_{12}}_{t_j} \mathcal{L}_{t_j} \Big[\sum\limits_{k=1}^{n_t} u_{t_k} \dfrac{\partial \mathcal{L}_{t_k}}{\partial y} + \sum\limits_{k=1}^{n_t} v_{t_k} \dfrac{\partial \mathcal{L}_{t_k}}{\partial x}+ C_2(r)\tan(\lambda)     \sum\limits_{k=1}^{n_t} {u}_{t_k} \mathcal{L}_{t_k}\Big] \Big)dA  \\
      &-\sum\limits_{i=1}^{2}\int_{\widehat{V}_e^i}      \dfrac{\partial}{\partial u_e}\Big( \sum\limits_{j=1}^{n_c} {\sigma_{12}}_{c_j} \mathcal{B}_{c_j} \Big[\sum\limits_{k=1}^{n_c} {u}_{c_k} \dfrac{\partial \mathcal{B}_{c_k}}{\partial y} + \sum\limits_{k=1}^{n_c}  v_{c_k} \dfrac{\partial \mathcal{B}_{c_k}}{\partial x}+ C_2(r) \tan(\lambda)     \sum\limits_{k=1}^{n_c} {u}_{c_k} \mathcal{B}_{c_j}\Big] \Big)dA .
\end{aligned}
 \end{equation}
 Applying the derivative to the above equation we have
   \begin{equation}
 \begin{aligned}\label{eq:pre_matrices}
      D_{u,e}^2  = &-\sum\limits_{i=1}^{2}\Big[\int_{T_e^i}     \sum\limits_{j=1}^{n_t} {\sigma_{12}}_{t_j} \mathcal{L}_{t_j} \Big(\dfrac{\partial \mathcal{L}_{\bar{e}}}{\partial y} + C_2(r)\tan(\lambda)    \mathcal{L}_{\bar{e}}\Big) dA  +\int_{\widehat{V}_e^i}  \sum\limits_{j=1}^{n_c} {\sigma_{12}}_{c_j} \mathcal{B}_{c_j} \Big(\dfrac{\partial \mathcal{B}_{\bar{e}}}{\partial y} + C_2(r) \tan(\lambda)      \mathcal{B}_{\bar{e}}\Big] \Big)dA,
\end{aligned}
 \end{equation}
 where again $\bar{e}$ refers again to the local index that corresponds to the global index of $e$. We define the matrices
   \begin{equation}\label{eq:matrices3}
 \begin{aligned}
     (\mathbb{N}^y_{t_e^i})_{j,k} &= \int_{T_e^i}  \mathcal{L}_{t_j}  \dfrac{\partial \mathcal{L}_{t_k}}{\partial y}  dA , \qquad j,k=\{1,\ldots,n_t\}, \quad i=1,2,\\
         (\mathbb{N}_{v_e^i}^y)_{j,k} &= \int_{\widehat{V}_e^i}  \mathcal{B}_{c_j}  \dfrac{\partial \mathcal{B}_{c_k}}{\partial y}  dA , \qquad j,k=\{1,\ldots,n_c\}, \quad i=1,2.
\end{aligned}     
 \end{equation}
  It follows from Eq. \eqref{eq:pre_matrices} that
 \begin{equation}
 \begin{aligned}
      D_{u,e}^2  = &-\sum\limits_{i=1}^{2}\Big[      \sum\limits_{j=1}^{n_t} {\sigma_{12}}_{t_j}\Big( (\mathbb{N}^y_{t_e^i})_{j,\bar{e}}+ C_2(r)\tan(\lambda)    (\mathbb{M}_{t_e^i})_{j,\bar{e}}\Big)   +  \sum\limits_{j=1}^{n_c} {\sigma_{12}}_{c_j}\Big( (\mathbb{N}^y_{v_e^i})_{j,\bar{e}} + C_2(r) \tan(\lambda)   (\mathbb{M}_{v_e^i})_{j,\bar{e}}\Big) \Big].
\end{aligned}
 \end{equation}
 For what concerns $D_{v,e}^2$, the procedure is similar until Eq. \eqref{eq:D2_der_ref}, where applying the derivative with respect to $v_e$ we obtain
    \begin{equation}
 \begin{aligned}\label{eq:pre_matrices2}
      D_{v,e}^2  = &-\sum\limits_{i=1}^{2}\Big[      \sum\limits_{j=1}^{n_t} {\sigma_{12}}_{t_j} (\mathbb{N}^x_{t_e^i})_{j,\bar{e}} +  \sum\limits_{j=1}^{n_c} {\sigma_{12}}_{c_j} (\mathbb{N}^x_{v_e^i})_{j,\bar{e}}  \Big].
\end{aligned}
 \end{equation}
 Last, let's consider $D_{u,e}^3$, hence the integral in Eq. \eqref{eq:D3}, which becomes
 \begin{equation}\label{eq:D3bis}
 D_{u,e}^3 = -\sum\limits_{i=1}^{2}\int_{T_e^i}\dfrac{\partial}{\partial u}\Big( \sigma_{22} \Big[ \dfrac{\partial v}{\partial y} + v \, C_3(r) \tan(\lambda)\Big] \Big) dA   -\sum\limits_{i=1}^{2}\int_{\widehat{V}_e^i} \dfrac{\partial}{\partial u}\Big( \sigma_{22} \Big[ \dfrac{\partial v}{\partial y} + v \, C_3(r) \tan(\lambda)\Big] \Big)  dA.
 \end{equation}
 Substituting the basis expansion we have
  \begin{equation} 
  \begin{aligned}
 D_{u,e}^3 = &-\sum\limits_{i=i}^{2}\int_{T_e^i} \dfrac{\partial}{\partial u_e}\Big(\sum\limits_{j=1}^{n_t}\sigma_{{22}_{t_j}} \mathcal{L}_{t_j} \Big[\sum\limits_{k=1}^{n_t}{v_{t_k}}\dfrac{\partial \mathcal{L}_{t_k}}{\partial y} +C_3(r) \tan(\lambda) \sum\limits_{k=1}^{n_t} {v}_{t_k} \mathcal{L}_{t_j}\Big]  \Big)dA \\
 &-\sum\limits_{i=1}^{2}\int_{\widehat{V}_e^i}\dfrac{\partial}{\partial u_e}\Big( \sum\limits_{j=1}^{n_c}\sigma_{{22}_{c_i}} \mathcal{B}_{c_j}\Big[ \sum\limits_{k=1}^{n_c}{v_{c_k}}\dfrac{\partial \mathcal{B}_{c_k}}{\partial y} +C_3(r) \tan(\lambda) \sum\limits_{k=1}^{n_c} {v}_{c_k} \mathcal{B}_{c_k}  \Big]\Big)dA.
 \end{aligned}
 \end{equation}
The application of the derivative gives $D_{u,e}^3 =0$.
For $D_{v,e}^3$ we have instead
     \begin{equation} \label{eq:D3_superFinal}
     \begin{aligned}
D_{v,e}^3 =   &-\sum_{i=1}^{2} \Big[\sum_{j=1}^{n_t}  {\sigma_{22}}_{t_j}\Big(  (\mathbb{N}^y_{t_e^i})_{j,\bar{e}}+C_3(r)\tan(\lambda_{e})(\mathbb{M}_{t_e^i})_{j,\bar{e}}\Big)+\sum_{j=1}^{n_c}  {\sigma_{22}}_{c_j}  \Big((\mathbb{N}^y_{v_e^i})_{j,\bar{e}}+C_3(r)\tan(\lambda_{e})(\mathbb{M}_{v_e^i})_{j,\bar{e}}\Big)\Big].
 \end{aligned}
 \end{equation}
 }
 
 \section*{Appendix B}
 \Giacomo{
Let $\Omega_e$ be the set of all edge points of the mesh and $\mathbf{y} \in \Omega_e$, then to test the accuracy of the approximation in Eq. \eqref{eq:gradApprox}, we consider the Taylor expansion of $f\in C^{2}(\Omega)$ at $\mathbf{x} \in \Omega_e$:
\begin{align}\label{eq:Taylor1}
     f(\mathbf{x}) &= f(\mathbf{y}) + (x_1 - y_1) \dfrac{\partial }{\partial x_1}f(\mathbf{y}) 
    + (x_2 - y_2) \dfrac{\partial }{\partial x_2}f(\mathbf{y}) \\
    &\qquad + \frac{1}{2}  \Big[  (x_1-y_1)^2 \frac{\partial^2}{\partial x_1^2}f(\mathbf{y}) 
+ (x_2-y_2) (x_1-y_1)\frac{\partial^2}{\partial x_1\partial x_2}f(\mathbf{y}) 
\\
& \qquad\qquad+(x_1-y_1) (x_2-y_2)\frac{\partial^2}{\partial x_2\partial x_1}f(\mathbf{y})
 +(x_2-y_2)^2\frac{\partial^2}{\partial x_2^2}f(\mathbf{y})\Big]
  + {\mathcal O}(\|\mathbf{x}-\mathbf{y}\|^{3}).
\end{align}
Let us define 
\begin{align}
g_1(\mathbf{x}) := 1,& \qquad g_2(\mathbf{x}): = (x_1-y_1), \qquad g_3(\mathbf{x}): = (x_2-y_2), \qquad
g_4(\mathbf{x}) := (x_1-y_1)^2,\\
\qquad g_5(\mathbf{x}) := &(x_2-y_2)(x_1-y_1), \qquad g_6(\mathbf{x})=g_5(\mathbf{x}), \qquad g_7(\mathbf{x}) = (x_2-y_2)^2,
\end{align}
which are all functions in $C^{2}(\Omega)$, and the coefficients
\begin{align}
c_1 := f(\mathbf{y}),& \qquad c_2:=\dfrac{\partial }{\partial x_1}f(\mathbf{y}), \qquad c_3:=\dfrac{\partial }{\partial x_2}f(\mathbf{y}) , \qquad c_4:=\dfrac{1}{2}\dfrac{\partial^2 }{\partial x_1^2}f(\mathbf{y}),\\
&c_5:=\dfrac{1}{2}\dfrac{\partial^2}{\partial x_1 \partial x_2}f(\mathbf{y}) , \qquad c_6:= c_5, \qquad c_7 = \dfrac{1}{2}\dfrac{\partial^2 }{\partial x_2^2}f(\mathbf{y}).
\end{align}
Then, given $\mathbf{y} \in \Omega_e$ and neglecting the third-order terms, $f$ can be approximated by a truncated Taylor expansion as
\begin{equation}
    f \approx \sum\limits_{i=1}^7 c_i g_i,
\end{equation}
with value at $\mathbf{x} \in \Omega_e$ approximated by
\begin{equation}
    f(\mathbf{x}) \approx \sum\limits_{i=1}^7 c_i g_i(\mathbf{x}).
\end{equation}
Note that $C^{2}(\Omega)$ is a linear space and $g_i \in C^{2}(\Omega)$ for $i=1,\ldots,7$ hence 
due to the linearity of $\mathcal{F}$ we have
\begin{equation}
    \mathcal{F}(f)(\mathbf{x}) \approx \sum\limits_{i=1}^7 c_i \mathcal{F}(g_i)(\mathbf{x}).
\end{equation}
Expanding the above sum we get
\begin{align}\label{eq:TaylorL}
     \mathcal{F}(f)(\mathbf{x}) &\approx f(\mathbf{y})\mathcal{F}(g_1)(\mathbf{x}) + \dfrac{\partial }{\partial x_1}f(\mathbf{y})\mathcal{F}(g_2)(\mathbf{x}) 
    +  \dfrac{\partial }{\partial x_2}f(\mathbf{y})\mathcal{F}(g_3)(\mathbf{x}) \\
    &\qquad + \frac{1}{2}  \Big[   \frac{\partial^2}{\partial x_1^2}f(\mathbf{y})\mathcal{F}(g_4)(\mathbf{x}) 
+ \frac{\partial^2}{\partial x_1\partial x_2}f(\mathbf{y})\mathcal{F}(g_5)(\mathbf{x}) 
\\
& \qquad\qquad+\frac{\partial^2}{\partial x_2\partial x_1}f(\mathbf{y})\mathcal{F}(g_6)(\mathbf{x})
 +\frac{\partial^2}{\partial x_2^2}f(\mathbf{y})\mathcal{F}(g_7)(\mathbf{x})\Big].
\end{align}
Because we want $\mathcal{F}$ to be an approximation of the gradient operator (ideally, we would like it to be exactly the gradient operator) on $\Omega_e$, we set the equality
\begin{align}
    \nabla f(\mathbf{x}) = \mathcal{F}(f)(\mathbf{x}), \quad \forall \mathbf{x} \in \Omega_e.
\end{align}
The above equality holds if the approximation below holds
\begin{equation}\label{eq:Lfinal}
\begin{aligned}
   \nabla f(\mathbf{x}) &\approx f(\mathbf{y})\mathcal{F}(g_1)(\mathbf{x}) + \dfrac{\partial }{\partial x_1}f(\mathbf{y})\mathcal{F}(g_2)(\mathbf{x}) 
    +  \dfrac{\partial }{\partial x_2}f(\mathbf{y})\mathcal{F}(g_3)(\mathbf{x}) \\
    &\qquad + \frac{1}{2}  \Big[   \frac{\partial^2}{\partial x_1^2}f(\mathbf{y})\mathcal{F}(g_4)(\mathbf{x}) 
+ \frac{\partial^2}{\partial x_1\partial x_2}f(\mathbf{y})\mathcal{F}(g_5)(\mathbf{x}) 
\\
& \qquad\qquad+\frac{\partial^2}{\partial x_2\partial x_1}f(\mathbf{y})\mathcal{F}(g_6)(\mathbf{x})
 +\frac{\partial^2}{\partial x_2^2}f(\mathbf{y})\mathcal{F}(g_7)(\mathbf{x})\Big].
\end{aligned}
\end{equation}
Hence, the relation above provides a sufficient condition for $\mathcal{F}$ to be a at least a second-order approximation of the derivative operator. In fact, if $\mathbf{x}=\mathbf{y}$, then \eqref{eq:Lfinal} implies
\begin{equation}
\begin{aligned}
     \nabla f(\mathbf{y}) &\approx f(\mathbf{y})\mathcal{F}(g_1)(\mathbf{y}) + \dfrac{\partial }{\partial x_1}f(\mathbf{y})\mathcal{F}(g_2)(\mathbf{y}) 
    +  \dfrac{\partial }{\partial x_2}f(\mathbf{y})\mathcal{F}(g_3)(\mathbf{y}) \\
    &\qquad + \frac{1}{2}  \Big[   \frac{\partial^2}{\partial x_1^2}f(\mathbf{y})\mathcal{F}(g_4)(\mathbf{y}) 
+ \frac{\partial^2}{\partial x_1\partial x_2}f(\mathbf{y})\mathcal{F}(g_5)(\mathbf{y}) 
\\
& \qquad\qquad+\frac{\partial^2}{\partial x_2\partial x_1}f(\mathbf{y})\mathcal{F}(g_6)(\mathbf{y})
 +\frac{\partial^2}{\partial x_2^2}f(\mathbf{y})\mathcal{F}(g_7)(\mathbf{y})\Big].
\end{aligned}
\end{equation}
The above relation shows that the following conditions are sufficient for $\mathcal{F}$ to be a at least a second-order approximation of $\nabla f$: 
\begin{equation}
    \begin{cases}
    \mathcal{F}(g_i)((x_e,y_e)) = [0,0]^T, i\neq 2, i \neq3, \\
    \mathcal{F}(g_2)((x_e,y_e)) = [1,0]^T, \\
    \mathcal{F}(g_3)((x_e,y_e)) = [0,1]^T.
    \end{cases},
\end{equation}
for any point $(x_e,y_e) \in \Omega_e$.
 }
 
\end{document}